\documentclass[11pt]{article}
\usepackage{amsmath, amssymb, amscd, amsthm, amsfonts}
\usepackage{graphicx}
\usepackage{hyperref}
\usepackage{graphicx}
\usepackage{booktabs}
\usepackage{multirow}
\usepackage{caption}
\usepackage{subfig}
\usepackage{graphicx}
\usepackage{subfig}
\usepackage{authblk}
\usepackage{algorithm}
\usepackage{algorithmic}
\usepackage{makecell}

\title{Adaptive Long-term Embedding with Denoising and Augmentation for Recommendation}
\author{
	Zahra Akhlaghi\thanks{zahra.akhlaghi@aut.ac.ir} \textsuperscript{1}
    and
	Mostafa Haghir Chehreghani\thanks{mostafa.chehreghani@aut.ac.ir (corresponding author)} \textsuperscript{1}\\
	Department of Computer Engineering, Amirkabir University of Technology (Tehran Polytechnic), Tehran, Iran}
\date{}

\begin{document}
\maketitle
\begin{abstract}
	The rapid growth of the internet has made personalized recommendation systems indispensable. Graph-based sequential recommendation systems, powered by Graph Neural Networks (GNNs), effectively capture complex user-item interactions but often face challenges such as noise and static representations. In this paper, we introduce the Adaptive Long-term Embedding with Denoising and Augmentation for Recommendation (ALDA4Rec) method, a novel model that constructs an item-item graph, filters noise through community detection, and enriches user-item interactions. Graph Convolutional Networks (GCNs) are then employed to learn short-term representations, while averaging, GRUs, and attention mechanisms are utilized to model long-term embeddings. An MLP-based adaptive weighting strategy is further incorporated to dynamically optimize long-term user preferences. Experiments conducted on four real-world datasets demonstrate that ALDA4Rec outperforms state-of-the-art baselines, delivering notable improvements in both accuracy and robustness. The source code is available at \url{https://github.com/zahraakhlaghi/ALDA4Rec}.
\\\\
\textbf{keyword:}
Sequential recommender systems, graph-based recommendation, graph neural networks, denoising and augmentation, short‑term and long‑term embeddings

\end{abstract}

	\section{Introduction}

The rapid expansion of the internet has led to an overwhelming volume of information, making it increasingly difficult for users to efficiently find relevant content \cite{wu2022survey, liao2022heterogeneous}. Recommender systems have emerged as a critical solution, leveraging user preferences and behavioral data to provide personalized suggestions, thereby improving user experience, engagement, and business outcomes \cite{zheng2021incorporating}. These systems are widely applied in domains such as e-commerce, social networks, and online shopping platforms. Various recommendation techniques have been explored, including content-based filtering, neural collaborative filtering \cite{CF}, and matrix factorization models \cite{MF}. While these approaches have demonstrated effectiveness, they often struggle to capture intricate dependencies and dynamically evolving user preferences. Recently, graph-based methods have gained significant attention due to their ability to model complex relationships between users and items, leading to more accurate and insightful recommendations \cite{he2020lightgcn, NGCF, UltraGCN, GC-MC}.

A crucial subfield of recommender system research is sequential recommendation, which aims to predict users' future interactions based on their historical behavior. Various methodologies have been developed for this task, including Markov-chain-based models \cite{Markov-chains}, recurrent neural networks (RNNs)-based models such as GRU4Rec \cite{GRU4Rec} and KrNN-P \cite{KrNN-P}, which capture sequential dependencies, and attention-based mechanisms that model item correlations, as seen in SASRec \cite{kang2018self} and TiSASRec \cite{li2020time}.
Recently, Graph Neural Networks (GNNs) \cite{DBLP:conf/iclr/KipfW17,zhou2020graph,DBLP:journals/natmi/Chehreghani22,gao2023survey,DBLP:journals/tjs/ZohrabiSC24,10.1145/3700790} have demonstrated significant potential in addressing various challenges and tasks on graph-structured data. Subsequently, they have been successfully applied to enhance sequential recommendation by effectively capturing user-item relationships over time. By leveraging graph structures, GNN-based models excel in representing complex item transitions within user sequences, leading to improved prediction accuracy and deeper insights into user behavior \cite{DGSR, SR-GNN, TEA}.

Despite significant advancements in graph-based sequential recommendation systems, several critical challenges persist, hindering their effectiveness in practical, real-world applications. \textbf{Firstly}, real-world user data inherently includes substantial noise, such as random clicks, temporary interests, and preferences that evolve over time \cite{Unbiased2020}. The definition and impact of this "noise" vary significantly across users; some exhibit diverse and shifting interests, whereas others maintain stable preferences. Existing noise-filtering techniques primarily address noise during the training phase \cite{VIB-GSL}, consequently increasing model complexity. Moreover, traditional methods typically lack strategies that proactively enhance recommendation accuracy---such as identifying and incorporating similar items---to reduce noise impact. \textbf{Secondly}, most conventional GNN-based recommendation models rely on static data representations, thereby limiting their ability to effectively adapt to users' dynamic and evolving preferences. Although recent studies have begun exploring dynamic and temporal graph networks, effectively modeling short-term user behaviors to accurately infer long-term interests remains a significant open challenge.

Among existing solutions, the model most closely related to ours is SelfGNN \cite{liu2024selfgnn}, which also partitions interactions into short‑term graphs and applies GNNs with a self‑supervised denoising loss. Nevertheless, SelfGNN has two inherent drawbacks: (i) its denoising relies on an auxiliary loss, which increases training complexity without augmenting the interaction graph; (ii) its long‑term embedding employs only GRU and attention, making it susceptible to vanishing gradients and the U‑shaped attention bias. Other recent models have combined diffusion processes with contrastive learning for denoising \cite{maskdiffusion,prototypelearning,latentdiffusion,smoothdiffusion,hierarchydiffusion}, but they target knowledge‑aware, multimodal, or social recommendation settings and do not address adaptive long‑term preference learning in a purely sequential GNN framework. Consequently, there exists a clear research gap: no existing approach integrates lightweight, preprocessing‑based graph denoising, graph augmentation with similar items, and an adaptive long‑term embedding that  mitigates U‑shaped bias while learning per‑user importance weights.

To address these limitations, we propose the Adaptive Long-term Embedding with Denoising and Augmentation for Recommendation (ALDA4Rec) method, a graph-based sequential recommendation model designed to enhance recommendation accuracy and robustness. Our model consists of three key components whose functional connections are as follows:
\begin{enumerate}
	\item 
	\textbf{Graph construction and denoising}: we segment user interactions into multiple time intervals and compute item similarities within each interval. Using these computed similarities, we construct an item-item graph where edges represent relationships between items. We then apply a community detection mechanism to identify and remove noisy data, simultaneously enriching the user's preference list with similar items. This preprocessing step directly supplies a cleaner and augmented interaction matrix to the next component, reducing the burden of noise removal from the learning phase.
	\item 
	\textbf{Long and short-term embedding learning and aggregation}: we generate a user-item graph for each time interval and use Graph Convolutional Networks (GCNs) to propagate higher-order collaborative signals for short-term embeddings. To capture long-term preferences, we combine mean pooling, GRU, and attention mechanisms. The mean pooling operates on the short‑term embeddings produced by the GCNs, providing a simple but effective baseline that counteracts the biases of the attention mechanism. The GRU and attention layers, in turn, model interval‑level and instant‑level dynamics. All three long‑term representations are then passed to the adaptive weighting module.
	\item 
	\textbf{Adaptive long-term embedding optimization}: the weights learned from the MLP dynamically determine the user-specific importance of long-term embeddings in predictions. We adaptively optimize each user's long-term embeddings based on their learned importance weights. This MLP takes the concatenation of mean‑level and interval‑level embeddings, producing a scalar weight per user. The weight controls the contribution of the GRU/attention based prediction versus the mean‑based prediction, thus personalizing the trade‑off between sequential dynamics and uniform temporal aggregation.
\end{enumerate}
The key contributions of our work can be summarized as follows:
\begin{itemize}
	\item We introduce a novel item similarity method and construct an item-item graph based on these similarities. Community detection is applied to each user's interacted items to filter noisy interactions and enrich user data with relevant items.
	\item We propose a graph-based recommendation framework that adapts to dynamic user interests through periodic collaborative learning, attentive sequential modeling, and averaged collaborative relationships.
	\item We use an MLP-based adaptive weighting mechanism to dynamically adjust the influence of long-term embeddings, ensuring personalized recommendations.
	\item We conduct extensive experiments on four real-world datasets, demonstrating that ALDA4Rec outperforms state-of-the-art baseline models. The experimental results validate both the effectiveness and practical applicability of our proposed approach.
\end{itemize}

The remainder of this paper is structured as follows: Section~\ref{sec:related} reviews the related work. Section~\ref{sec:preliminaries} introduces the preliminaries and key definitions. Section~\ref{sec:methodology} elaborates on our proposed method. Section~\ref{sec:experiments} presents the experimental results and evaluates the model's performance. Finally, Section~\ref{sec:conclusion} concludes the paper and highlights future research directions.

\section{Related work}
\label{sec:related}

In this section, we review recent advancements in three closely related fields: graph-based recommendation, sequential recommender systems, and denoising in recommender systems.

\subsection{Graph-based recommendation}
Recent advancements in GNNs lead to the development of various models for recommender systems. GC-MC \cite{GC-MC} applies a graph-based autoencoder for matrix completion, while NGCF \cite{NGCF} captures high-order connectivity by propagating embeddings in the user-item graph. GCCF \cite{GCCF} eliminates non-linearities and introduces a residual structure to address over-smoothing in sparse interactions. LightGCN \cite{he2020lightgcn} simplifies GCN-based collaborative filtering by removing weight transformations and nonlinear activations. UltraGCN \cite{UltraGCN} approximates infinite-layer graph convolutions with constraint loss, enabling flexible edge weight assignments.
HetroFair \cite{HetroFair} introduces fairness-aware embeddings by employing fairness-aware attention to mitigate node degree effects and heterophily feature weighting to assign distinct importance to different features.
Gholinejad and Haghir Chehreghani \cite{gholinejad2025disentanglingpopularityqualityedge}
propose a GNN-based recommendation model that disentangles items' popularity and quality. It introduces an edge classification technique to differentiate between popularity bias and genuine quality disparities among items, and uses cost-sensitive learning to adjust the misclassification penalties.
PBiLoss \cite{pbiloss} targets fairness in graph-based collaborative filtering by introducing a popularity-aware bilateral loss. The method adds a regularization term that penalizes predictions that systematically over-rank popular items while under-ranking niche ones.

Recent advancements in temporal GNNs have significantly enhanced sequential recommendation tasks, addressing key challenges such as capturing sequential dependencies and dynamic user-item interactions. SR-GNN \cite{SR-GNN} models session sequences as graphs to capture item transitions. TEA \cite{TEA} leverages dynamic heterogeneous graphs for user-item interactions. DGSR \cite{DGSR} focuses on individual user sequences but overlooks dynamic collaborative signals. Redrec \cite{Redrec} incorporates relation-aware GCNs with time decay functions to model item relationships.  SelfGNN \cite{liu2024selfgnn} encodes short-term graphs using time intervals and employs GNNs to model short-term collaborations.
It is the most similar work to our method (ALDA4Rec).
Quantitatively, ALDA4Rec outperforms SelfGNN by an average of 7.5\% in HR@10 and 7.1\% in NDCG@10 across the four datasets (see Table~\ref{tab:results}). The inherent drawbacks of SelfGNN that our method addresses are: (i) its denoising is performed via a self‑supervised loss, which adds training complexity and does not augment the graph; (ii) its long‑term embedding uses only GRU and attention, which are prone to the U‑shaped bias and vanishing gradients; (iii) it lacks an adaptive mechanism to balance different types of long‑term representations per user. ALDA4Rec resolves these by preprocessing denoising and augmentation, introducing mean‑level embeddings, and employing an MLP‑based adaptive weighting.

\subsection{Sequential recommender systems}
Sequential recommender systems play a vital role in analyzing user interaction patterns over time, enabling personalized recommendations. These patterns provide key insights into both short-term user interests and the gradual evolution of their preferences. Traditional approaches based on RNNs demonstrate effectiveness in modeling sequential dependencies. GRU4Rec \cite{GRU4Rec} introduces an RNN-based session-based recommendation model, while KrNN-P \cite{KrNN-P} incorporates neighboring sequences into RNNs to capture local relationships, proposing the K-plet RNN (KrNN) to jointly model multiple sequences. However, RNNs struggle with long-term dependencies, leading to the adoption of self-attention mechanisms. SASRec \cite{kang2018self} leverages self-attention to capture long-term user preferences efficiently, while TiSASRec \cite{li2020time} extends this by incorporating absolute positions and time intervals between interactions. BERT4Rec \cite{BERT4Rec} further improves recommendation accuracy by utilizing deep bidirectional self-attention, allowing each item in a user's history to contextualize information from both preceding and succeeding interactions.

Recent models effectively combine recurrent neural networks (RNNs) and self-attention mechanisms for sequential recommendations. SelfGNN \cite{liu2024selfgnn} enhances user-item representations by applying attention mechanisms to GRU hidden states and embedding sequences, capturing detailed user preferences. Contrastive learning techniques also improve interest modeling. For example, CLSR \cite{CLSR} utilizes a dual-encoder to capture long-term and short-term interests, refining embeddings via contrastive learning and dynamically aggregating them with attention. These developments emphasize the importance of accurate and adaptive sequential recommendation systems.

\subsection{Denoising in recommender systems}
Recommender systems depend heavily on accurate historical interactions, but real-world data often includes noise from clickbait or accidental interactions, degrading performance. Several methods address this issue.
NiDen \cite{NiDen} leverages neighborhood data and dynamic thresholding for noise identification and mitigation. LLM4DSR \cite{LLM4DSR} uses large language models (LLMs) to detect noisy items through self-supervised fine-tuning. SelfGNN \cite{liu2024selfgnn} corrects short-term interaction graphs based on long-term user interests. GraphDA \cite{fan2023graphda} enhances user-item interaction matrices with top-K sampling and augmented correlations. UDT \cite{UDT} distinguishes user behaviors into willingness and action phases to capture distinct noise patterns. VIB-GSL \cite{VIB-GSL} applies a variational information bottleneck to extract informative graph structures. SubGCL \cite{SubGCL} employs perceptual signal extraction, self-attention mechanisms, and contrastive learning for robust denoising in implicit feedback. DCF \cite{10.1145/3637528.3671692} introduces a double correction framework that stabilizes noise detection using temporal loss patterns, distinguishes hard samples from noise, and applies progressive label correction to reduce data sparsity.

Despite their effectiveness, existing denoising strategies share several common limitations: (1) Most methods incorporate noise detection as part of the training process (e.g., via self‑supervised losses or auxiliary modules), which increases model complexity and training time. (2) They rarely augment the interaction graph with inferred relevant edges, missing an opportunity to compensate for sparse or missing interactions. (3) Many approaches rely on heuristic thresholds (e.g., top‑K sampling or similarity cutoffs) that are sensitive to dataset characteristics and require extensive tuning. (4) The majority of denoising techniques are evaluated only on accuracy metrics without analyzing their impact on the temporal bias of attention mechanisms. Our preprocessing method addresses these limitations by performing denoising and augmentation offline, using community detection that adapts to each user's item subgraph, and by explicitly showing how the mean‑level embedding mitigates attention bias.

\section{Preliminaries}
\label{sec:preliminaries}

Let \( U = \{u_1, u_2, \dots, u_I\} \) represents the user set, where \( |U| = I \), and let \( V = \{v_1, v_2, \dots, v_J\} \) denotes the item set, where \( |V| = J \). To capture the temporal dynamics of interactions, we divide the time range from \( t_a \) (first recorded interaction) to \( t_b \) (latest recorded interaction) into \( T \) equal-length intervals, where \( T \) is a configurable parameter. The duration of each interval is given by \( (t_b - t_a) / T \). At each time step \( t \), an adjacency matrix \( A_t \in \mathbb{R}^{I \times J} \) encodes the implicit user-item interactions. Specifically, if user \( u_i \) engages with item \( v_j \) during time step \( t \), then \( A_{t,i,j} = 1 \); otherwise, it remains 0.

To construct an item-item relationship graph, we leverage user interaction sequences within each interval. If a user first selects item \( v_a \) and later interacts with item \( v_b \), an undirected edge is formed between them, with its weight increasing by one. The strength of each edge reflects the frequency of co-occurrence between item pairs. These weights are subsequently normalized to range between 0 and 1. The resulting adjacency matrix \( Z_t \in \mathbb{R}^{J \times J} \) represents the item-item graph at time step \( t \).

The central task is to predict user-item interactions in the future based on past interaction patterns. Given a sequence of adjacency matrices \( \{A_t \mid 1 \leq t \leq T \} \), our objective is to estimate the interactions in the next time interval, denoted as \( \hat{A}_{T+1} \). This task is framed as an optimization problem, where we seek to minimize the reconstruction loss \( \mathcal{L}_{rec} \) between the predicted interactions \( \hat{A}_{T+1} \) and the ground truth \( A_{T+1} \):

\begin{equation}
	\hat{A}_{T+1} = \arg\min_{\hat{A}} \mathcal{L}_{rec} (\hat{A}, A_{T+1}).
\end{equation}

\section{Methodology}
\label{sec:methodology}

\begin{figure*}[h]
	\centering
	\includegraphics[width=0.9\textwidth]{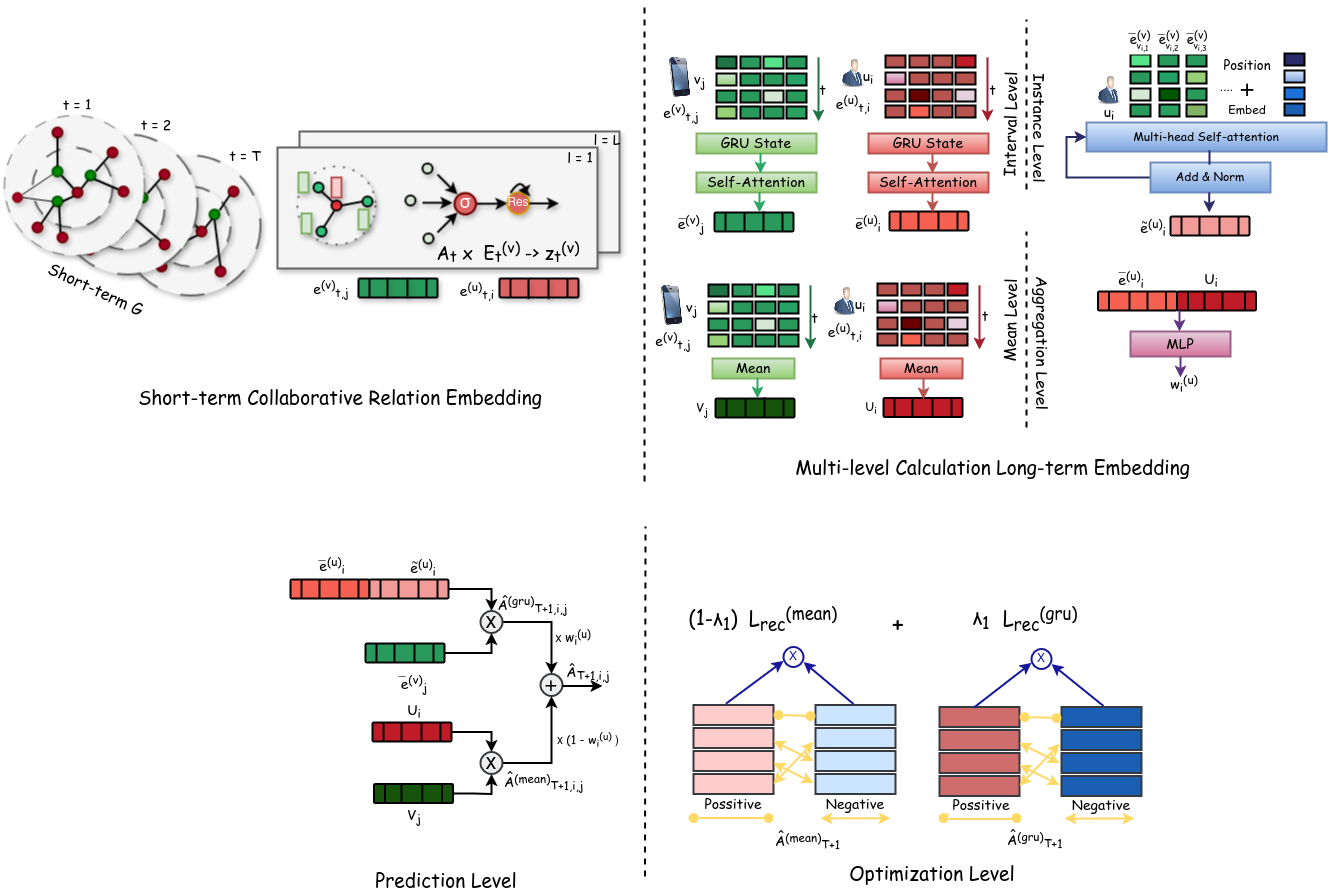}
	\caption{Overall framework of the proposed ALDA4Rec model}
	\label{fig:frame}
\end{figure*}

In this section, we first present our preprocessing method  to refine $A_t$.
Following that, we present different components of our proposed ALDA4Rec framework.
The overall architecture of the ALDA4Rec model is depicted in Figure \ref{fig:frame}.

\subsection{Preprocessing}
In this section, we describe the preprocessing method used to enhance user interactions in the matrix \(\{ A_t \mid 1 \leq t \leq T \}\).
In our method, first we compute similarity scores between items.
Then, based on the computed similarities, we identify and remove noisy data using a community detection method,
and add new interactions for users to augment the user-item graph.

\subsubsection{Similarity-based graph construction}
We consider two key factors when calculating the similarity between items. First, some items are highly popular among users and frequently appear in most shopping baskets. As a result, they are more likely to be interacted with after other items, even if they are not strongly related. Second, some items may indeed be similar, but due to the time gap between a user's visits to the shopping basket, this similarity may not be accurately captured. 

While traditional similarity measures such as cosine or Jaccard similarity are commonly used in recommender systems, they are not well-suited to address these challenges. Specifically, cosine similarity is sensitive to item popularity and may overestimate similarity between frequently co-occurring items, regardless of their contextual relevance. Jaccard similarity, which focuses on set overlap, also does not incorporate temporal or directional information and fails to distinguish between co-occurrence due to popularity and that due to true relevance. To address these limitations, we use the following equation to compute the similarity between items:
\begin{equation}
	I_{t}^{\text{sim}}[i,j] = \frac{(Z_t + I) \cdot (Z_t + I)^T [i,j]}{Z_t[i] \cdot Z_t[j]^T},
	\label{eq:similarity}
\end{equation}
where \( I \) is the identity matrix. The numerator \((Z_t + I)(Z_t + I)^T\) expands the connectivity by considering paths of up to length two: the identity matrix adds self-loops, so that each item's direct co‑occurrence (paths of length 1) is counted with an extra weight of 1, while the term \(Z_t Z_t^T\) captures two‑step paths (items co‑occurring through a common neighbor). This design gives higher importance to direct co‑occurrences than to indirect ones, which is desirable because direct transitions reflect stronger associations. The denominator \(Z_t[i] \cdot Z_t[j]^T\) is the product of the total co‑occurrence counts (popularity) of items \(i\) and \(j\). Dividing by this product reduces the bias toward popular items that frequently co‑occur with many others regardless of true similarity. The resulting score \(I_{t}^{\text{sim}}[i,j]\) is not necessarily symmetric, which allows capturing directional preferences (e.g., item \(i\) often precedes \(j\) but not vice versa). These similarity scores serve as the basis for the next steps in preprocessing: identifying and down-weighting noisy interactions, and augmenting the interaction graph with inferred user-item links.

\subsubsection{Noise detection}
\begin{figure}[h]
	\centering
	\includegraphics[width=0.4\textwidth]{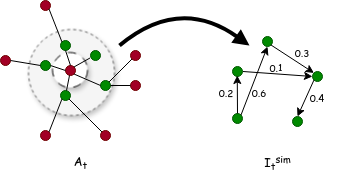}
	\caption{Extracting user interactions from \( I_t^{sim} \)}
	\label{fig:sim1}
\end{figure}

To identify noise, we first extract the items a user interacts with during a specific time period. We then apply the Louvain community detection algorithm \cite{louvain_community} to the corresponding subset of the \( I_t^{\text{sim}} \) graph that represents these items (Figure~\ref{fig:sim1}). The Louvain algorithm is a widely used method for detecting communities in large networks by optimizing modularity---a measure that compares the density of links within communities to the density of links between communities.

This process is performed independently for each user and each time period. Specifically, for every user at each time step \( t \), we construct a user-specific subgraph whose nodes are the items the user interacted with, and whose edge weights are given by the corresponding similarity scores from \( I_t^{\text{sim}} \). We then apply the Louvain algorithm to this subgraph to identify groups of closely related items. Forming item communities in this personalized manner allows the method to adapt to variations in user interests and behavior. For example, if a user is only interested in music, we treat sports-related items as noise, whereas for another user, both music and sports items may be considered relevant.

Finally, we identify items that do not belong to any community as noise since the user is not interested in their similar items, and they remain ungrouped. We then reduce their associated edge weights in the matrix \( A_t \), with the degree of reduction determined by a hyperparameter. This adjustment minimizes the influence of noisy items on subsequent analyses, enhancing the accuracy of processing relevant data.
\subsubsection{Edge augmentation}
To augment new edges, we classify users into two groups. The first group includes previously active users who have since become inactive. For these users, we generate interactions with items similar to those they previously engaged with, simulating their potential behavior. Specifically, if a user was inactive at time \( t \) but interacted with item \( i \) at \( t-1 \), and item \( i \) is sufficiently similar to item \( j \) at \( t \), we add an interaction with \( j \) to \( A_t \). The second group consists of active users at time \( t \). To enrich their interactions, we incorporate similar but ignored items. If item similarity at \( t \) exceeds a threshold and the user interacted with \( i \), an interaction with \( j \) is added in proportion to their similarity.

\subsection{Embedding learning and prediction}
This section explores our method of learning user and item embedding matrices from short-term graphs and utilizing them for prediction.
\subsubsection{Short-term embedding learning}
To capture short-term patterns, we construct embedding matrices using short-term interaction graphs, following an approach similar to SelfGNN \cite{liu2024selfgnn}. Each user \( u_i \) and item \( v_j \) are mapped to a \( d \)-dimensional latent space, with their embeddings at time \( t \) represented as \( e_{t,i}^{u} \) and \( e_{t,j}^{v} \), respectively. These embeddings are organized into matrices \( E_t^{(u)} \in R^{I \times d} \) for users and \( E_t^{(v)} \in R^{J \times d} \) for items. Inspired by LightGCN \cite{he2020lightgcn}, we adopt a simplified GCN for modeling short-term interactions:
\begin{equation}
	z_{t,i}^{(u)} = \sigma(A_t \cdot E_t^{(u)}), \quad z_{t,j}^{(v)} = \sigma(A_t \cdot E_t^{(v)})
\end{equation}
where \( z_{t,i}^{(u)} \) and \( z_{t,j}^{(v)} \) aggregate information from neighboring nodes, and \( \sigma \) denotes the LeakyReLU activation function. To mitigate overfitting, we apply edge dropout randomly and use deeper GCN layers for enhanced information propagation.  

At the \( l \)-th layer, the embeddings are updated via message passing as follows:
\begin{equation}
	e_{t,i,l}^{(u)} = z_{t,i,l}^{(u)} + e_{t,i,l-1}^{(u)}, \quad e_{t,j,l}^{(v)} = z_{t,j,l}^{(v)} + e_{t,j,l-1}^{(v)}
\end{equation}
To construct the final embedding representation, we concatenate embeddings from all layers:
\begin{equation}
	e_{t,i}^{(u)} = e_{t,i,1}^{(u)} || \cdots || e_{t,i,L}^{(u)}, \quad e_{t,j}^{(v)} = e_{t,j,1}^{(v)} || \cdots || e_{t,j,L}^{(v)}
\end{equation}

\subsubsection{Long-term embedding learning}
We learn long-term embeddings through three phases: interval-level, instant-level, and mean-level.
In the following, we describe each of these phases in details.

\paragraph{Interval-level long-term embedding}
We generate long-term interval-level embeddings similar to the SelfGNN \cite{liu2024selfgnn} model. As we described in the previous section, we extract short-term embeddings from graphs corresponding to each time period for both users and items. We then arrange these embeddings into sequences and process them using a Gated Recurrent Unit (GRU) network. The GRU updates its hidden state at each time step, allowing us to capture temporal dependencies in the data. We compute the hidden state at time \(t\) as follows:
\begin{equation}
	h^{(u)}_{t,i} = GRU(e^{(u)}_{t,i}, h^{(u)}_{t-1,i}), \quad h^{(v)}_{t,j} = GRU(e^{(v)}_{t,j}, h^{(v)}_{t-1,j})
\end{equation}
In this equation, \(h^{(u)}_{t,i} \in R^d\) and \(h^{(v)}_{t,j} \in R^d\) represent the GRU's hidden states, which encode sequential patterns over time. We define the sequences for users \(u_i\) and items \(v_j\) as:
\begin{equation}
	S^{\text{intv}}_i = (h^{(u)}_{1,i}, \dots, h^{(u)}_{t,i}, \dots, h^{(u)}_{T,i}),
	\quad S^{\text{intv}}_j = (h^{(v)}_{1,j}, \dots, h^{(v)}_{t,j}, \dots, h^{(v)}_{T,j})
\end{equation}
Next, we apply a multi-head dot-product attention, denoted as \(\text{Self-Att}(\cdot)\), to extract meaningful temporal features from the sequences \(S^{\text{intv}}_i\) and \(S^{\text{intv}}_j\):
\begin{equation}
	\bar{H}^{(u)}_i = \text{Self-Att}(S^{\text{intv}}_i), \quad \bar{H}^{(v)}_j = \text{Self-Att}(S^{\text{intv}}_j)
\end{equation}
Finally, we aggregate the extracted features over all time steps to obtain the long-term embeddings:
\begin{equation}
	\bar{e}^{(u)}_i = \sum_{t=1}^{T} \bar{H}^{(u)}_{i,t}, \quad \bar{e}^{(v)}_j = \sum_{t=1}^{T} \bar{H}^{(v)}_{j,t}
\end{equation}
Here, \(\bar{e}^{(u)}_i\) and \(\bar{e}^{(v)}_j\) represent the final long-term embedding vectors, which capture the temporal evolution of user and item interactions over the given intervals.

\paragraph{Instant-level long-term embedding}
Similar to the SelfGNN \cite{liu2024selfgnn} model, we apply the self-attention mechanism to sequences of items that users have interacted with. We denote the \( m \)-th interacted item of user \( u_i \) as \( v_{i,m} \), where \( m \in \{1, 2, ..., M\} \) and \( M \) represents the maximum length of interactions. We represent the sequence of user \( u_i \)'s actions as:  
\begin{equation}
	S^{\text{inst}}_{i,0} = (\bar{e}^{(v)}_{v_{i,1}} + \mathbf{p}_1, \dots, \bar{e}^{(v)}_{v_{i,M}} + \mathbf{p}_M) 
\end{equation}
Here, we define \( \bar{e}^{(v)}_{v_{i,m}} \in R^d \) as the interval-level long-term embedding of item \( v_{i,m} \), while \( \mathbf{p}_m \in R^d \) is a learnable parameter that captures the positional significance of \( m \). To extract sequential patterns, we apply \( L_a \) layers of self-attention with residual connections:  
\begin{equation}
	S^{\text{inst}}_{i,l} = \sigma(\text{Self-Att}(S^{\text{inst}}_{i,l-1})) + S^{\text{inst}}_{i,l-1}, \quad \tilde{e}^{(u)}_i = \sum S^{\text{inst}}_{i,L_a} 
\end{equation}
In the equation above, \( S^{\text{inst}}_{i,l} \) denotes the sequence of user \( u_i \) at the \( l \)-th iteration of self-attention. We use the vector \( \tilde{e}^{(u)}_i \in R^d \) to capture the sequential dependencies within the user's interactions, thereby modeling the user's preferences.  

\paragraph{Mean-level long-term embedding}
GRU neural networks, commonly used for learning embeddings, process sequential data by iteratively updating their hidden states at each time step. However, like other recurrent architectures, they suffer from the vanishing gradient problem, limiting their ability to capture long-range dependencies. Additionally, attention mechanisms exhibit the U-shaped attention distribution, which disproportionately emphasizes tokens at the beginning and end of sequences \cite{barbero2024transformers, DBLP:journals/corr/abs-2411-15671}. To mitigate these challenges and improve prediction performance in the SelfGNN \cite{liu2024selfgnn} model--where GRUs with attention mechanisms are employed for long-term embedding extraction--we propose an alternative approach. Instead of relying solely on GRU-based embeddings, we compute long-term embeddings by averaging those derived from short-term graphs. Formally, this process is defined as:
\begin{equation}
	\begin{split}
		U_i &= \text{mean}\left(e^{(u)}_{1,i}, e^{(u)}_{2,i}, \ldots, e^{(u)}_{T,i}\right) \\[2pt]
		V_j &= \text{mean}\left(e^{(v)}_{1,j}, e^{(v)}_{2,j}, \ldots, e^{(v)}_{T,j}\right) 
	\end{split}\label{eq:mean-long-embed}
\end{equation}
where \( U_i \in \mathbb{R}^d \) and \( V_j \in \mathbb{R}^d \) represent the mean embeddings over short-term graphs for user \( u_i \) and item \( v_j \), respectively, computed across \( T \) time steps. This uniform mean aggregation treats each time step equally, mitigating the positional bias introduced by attention mechanisms. By smoothing the contribution of all sequence elements, it reduces the U-shaped bias and enhances the representation of mid-sequence information.

\begin{figure}[h]
	\centering
	{
		\includegraphics[scale=.18]{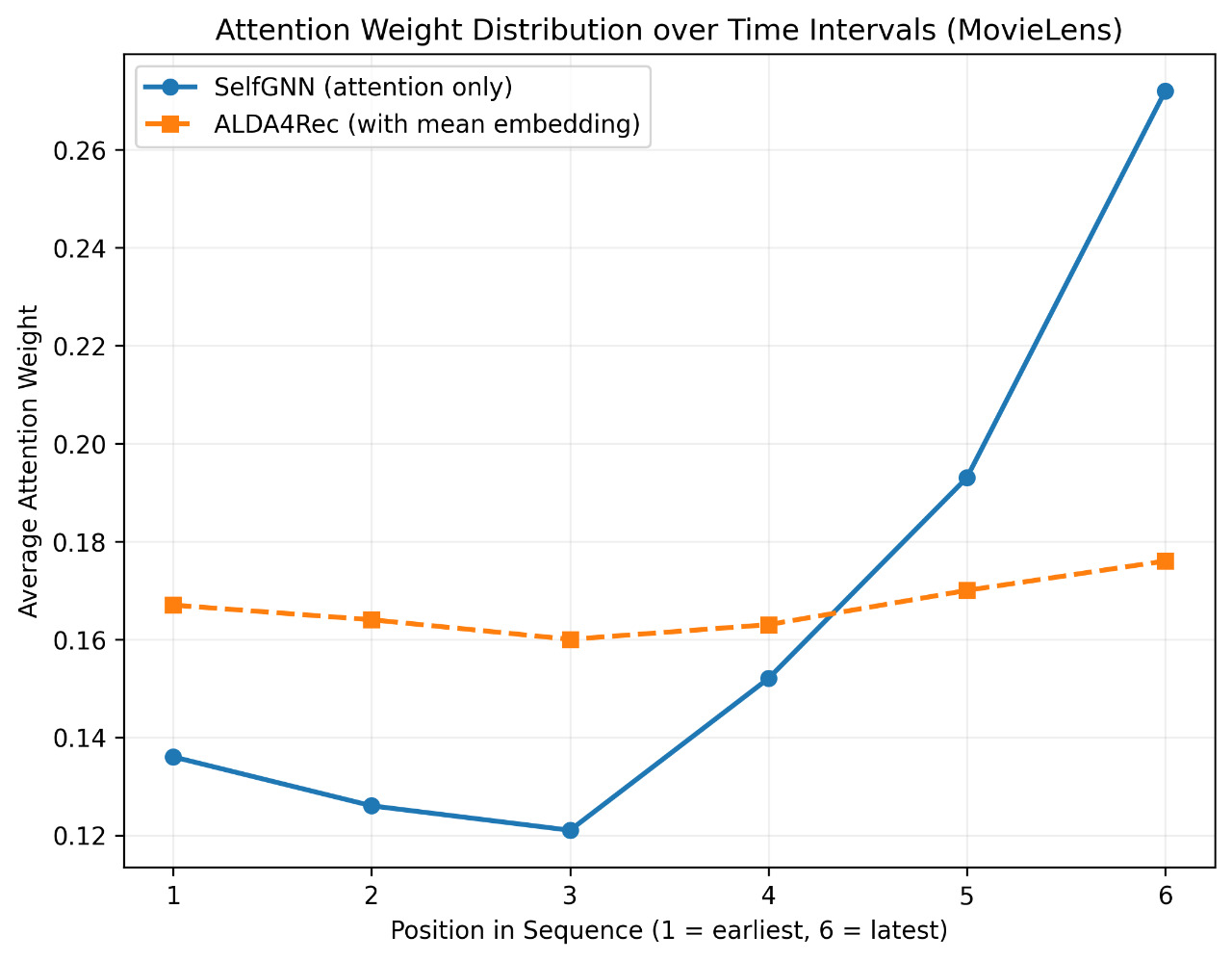}
		
	}
	\caption{Position‑wise attention weights on the Movielens dataset (6 time intervals). SelfGNN exhibits a U‑shaped pattern, whereas ALDA4Rec produces a nearly uniform distribution due to the contribution of mean‑level embeddings. This provides direct evidence that the U‑shaped bias is alleviated.}
	\label{fig:attention_bias}	
\end{figure}

To empirically examine whether mean-level embeddings alleviate the U-shaped attention bias, we conduct a position-wise analysis of the attention weights generated by the interval-level self-attention module in both SelfGNN and our proposed model with mean-level embeddings. For a fair comparison, the attention weights are L2-normalized across positions. Figure~\ref{fig:attention_bias} presents the average attention weight assigned to each position in the sequence of time intervals (with sequence length \(T=6\) on the Movielens dataset). In SelfGNN, the attention distribution exhibits a clear U-shaped pattern: weights are relatively high at the earliest position, decrease across subsequent positions, and then increase again toward the most recent position, resulting in lower emphasis on the middle intervals. This behavior reflects a positional bias that underutilizes information from intermediate time steps.	After incorporating the mean-level embedding and adaptively combining it with the attention output via the learned MLP weight \(w_i\), the resulting attention distribution becomes more balanced across positions. This effect can be attributed to the mean embedding, which provides a global summary signal and contributes uniformly across all intervals. Overall, the observed flattening of the attention distribution suggests that the proposed approach mitigates positional bias. This more balanced utilization of temporal information is consistent with the improvements in recommendation performance reported in Table~\ref{table:comparison-test1}.

\subsubsection{Embedding aggregation and prediction}

Prior to making predictions, we employ an MLP to assess the significance of each long-term embedding for individual users.
\begin{equation}
	\begin{split}
		h^{(u)}_i &= \text{MLP}\left([U_i, \bar{e}^{(u)}_i]\right) \\[2pt]
		w_i &= \sigma\left(h^{(u)}_i\right) \label{ef:weight}
	\end{split}
\end{equation}

In Equation \ref{ef:weight}, we concatenate the mean-level and interval-level long-term embeddings and feed them into an MLP network. The final layer employs a sigmoid activation function ($\sigma$), and we interpret $w_i$ as the importance weight of sequential long-term embeddings across time intervals for user $u_i$. 

Unlike standard attention mechanisms---which typically compute normalized weights across a set of embeddings based on pairwise similarity (e.g., using softmax over query-key scores)---our MLP-based approach produces an \textit{independent scalar weight} for each user. This weight is derived directly from the concatenated embedding pair $[U_i, \bar{e}^{(u)}_i]$, enabling the model to \textit{learn a global blending preference} between the two types of long-term representations on a per-user basis, without relying on alignment or distribution-based normalization across time steps.

We compute predictions based on the obtained long-term embeddings using the following equations:  \begin{equation} \begin{aligned} \hat{A}^{(\text{mean})}_{T+1,i,j} &= U_i \cdot V_j, \ \hat{A}^{(\text{gru})}_{T+1,i,j} &= (\bar{e}^{(u)}_i + \tilde{e}^{(u)}_i) \cdot \bar{e}^{(v)}_j. \end{aligned} \end{equation}
Here, we denote the prediction of user $u_i$'s interaction with item $v_j$ in the $(T+1)$-th time slot as $\hat{A}^{(\text{mean})}_{T+1,i,j}$, which is derived from the mean-level long-term embedding. Meanwhile, $\hat{A}^{(\text{gru})}_{T+1,i,j}$ represents the prediction obtained from long-term embeddings via the GRU neural network and attention mechanism, integrating both interval-level and instant-level embeddings.

The weight computed in Equation \ref{ef:weight} serves as a coefficient for combining these predictions. We formulate the final prediction of user $u_i$'s interaction with item $v_j$ at time $T+1$ as: 
\begin{equation*}
	\hat{A}_{T+1,i,j} = w_i \cdot \hat{A}^{(gru)}_{T+1,i,j} + (1 - w_i) \cdot \hat{A}^{(\text{mean})}_{T+1,i,j}. \label{eq:final-pred}
\end{equation*}
This formulation enhances the model's flexibility, allowing us to optimally integrate information for improved future interaction predictions.

\subsection{Loss functions}
In our proposed method, we use the coefficient obtained from Equation \ref{ef:weight} to optimize the model. Our approach adjusts the embeddings based on the weight assigned to each by the user. Accordingly, we apply the following equation to optimize the recommendations:
\begin{equation*} 
	\begin{split} 
		\mathcal{L}_{rec}^{(gru)} = \sum_{i=1}^{I} \sum_{k=1}^{N_{pr}} \max \left(0, 1 - \left( w_i \cdot (\hat{A}_{T+1,i,p_k}^{(gru)} - \hat{A}_{T+1,i,n_k}^{(gru)}) \right) \right)	\\[2pt]
		\mathcal{L}_{rec}^{(mean)} = \sum_{i=1}^{I} \sum_{k=1}^{N_{pr}} \max \left(0, 1 - \left( (1-w_i) \cdot (\hat{A}_{T+1,i,p_k}^{(mean)} - \hat{A}_{T+1,i,n_k}^{(mean)}) \right) \right)
		\label{eq:final-predict}
	\end{split}
\end{equation*}
We utilize \(\mathcal{L}_{rec}^{(gru)}\) to optimize the long-term embeddings extracted by the GRU network, while \(\mathcal{L}_{rec}^{(mean)}\) optimizes the embeddings computed through averaging. Here, \( N_{pr} \) denotes the number of samples, and \( p_k \) and \( n_k \) represent the indices of the \( k \)-th positive and negative samples, respectively. We define the final optimization equation as follows:
\begin{equation} 
	\mathcal{L} = \lambda_1 \cdot \mathcal{L}^{(gru)}_{rec} + (1- \lambda_1) \cdot \mathcal{L}^{(mean)}_{rec} + \lambda_2 \cdot \|\Theta\|_F^2 
\end{equation}
In this equation, the coefficients \(\lambda_1\) and \(\lambda_2\) serve as hyperparameters to balance the importance of different terms in the optimization function. The term \(\|\Theta\|_F^2\) represents the weight decay of L2 regularization, where \(\lambda_2\) controls its influence. Finally, \(\mathcal{L}\) denotes the overall loss function that we use for model optimization.
\subsection{The algorithm}
Algorithm \ref{alg:alg1} presents the pseudocode of the proposed ALDA4Rec model. As described earlier, the process begins by generating user and item embeddings for each short-term graph (lines 2--5). Following this, two types of long-term embeddings are extracted: the first is computed using a GRU combined with an attention mechanism to capture interval-level and instance-level long-term dependencies (lines 6--8), while the second is obtained by averaging the embeddings from the short-term graphs (line 9). To further refine these representations, an MLP is employed to assess the significance of the long-term embeddings for each user (line 10). Finally, these embeddings are utilized for prediction, and throughout the process, the loss function is continuously optimized to enhance the model's performance.
\begin{algorithm}[!h]
	\caption{Inference Process of the ALDA4Rec Framework.\label{alg:alg1}}
	\begin{algorithmic}[1]
		\REQUIRE Pre-processed graph sequences $\{A_t \mid 1 \leq t \leq T\}$, number of samples $N$, maximum epochs $E$, loss function weights $\lambda_1$, regularization coefficient $\lambda_2$, and learning rate $\rho$
		
		\ENSURE Optimized parameter set $\Theta$
		
		\STATE Initialize $\Theta$ with random values
		\FOR{epoch $e = 1$ to $E$}
		\FOR{each time step $t = 1$ to $T$}
		\STATE Extract user short-term embeddings $e^{(u)}_{t,i}$ and item short-term embeddings $e^{(v)}_{t,j}$ for time $t$
		\ENDFOR
		\STATE Construct feature evolution sequences $S^{\text{intv}}_{u_i}$ for users and $S^{\text{intv}}_{v_j}$ for items across intervals
		\STATE Compute interval-based long-term user embeddings $\bar{e}^{(u)}_i$ and item embeddings $\bar{e}^{(v)}_j$
		\STATE Capture instance-level long-term sequential user features $\tilde{e}^{(u)}_i$
		\STATE Compute temporal average representations for users ($U_i$) and items ($V_j$)
		\STATE Concatenate $\bar{e}^{(u)}_i$ with $U_i$ and use an MLP to determine weighting factors for long-term representation in the optimization process
		\STATE Sample a mini-batch $U$ containing $N_{pr}$ positive-negative user-item pairs
		\STATE Initialize loss: $\mathcal{L} = \lambda_2 \|\Theta\|^2_F$
		\FOR{each user $u_i$ in $U$}
		\STATE Compute predictions $\hat{A}^{(mean)}_{T+1,i,p_k}$, $\hat{A}^{(mean)}_{T+1,i,n_k}$, $\hat{A}^{(gru)}_{T+1,i,p_k}$, and $\hat{A}^{(gru)}_{T+1,i,n_k}$
		\STATE Update $\mathcal{L}$ with hinge loss components:
		\STATE $\mathcal{L} \mathrel{+}= \sum_{k=1}^{N_{pr}} \max(0, 1 - (w_i \cdot (\hat{A}^{(gru)}_{T+1,i,p_k} - \hat{A}^{(gru)}_{T+1,i,n_k})))$
		\STATE $\mathcal{L} \mathrel{+}= \sum_{k=1}^{N_{pr}} \max(0, 1 - ((1-w_i) \cdot (\hat{A}^{(mean)}_{T+1,i,p_k} - \hat{A}^{(mean)}_{T+1,i,n_k})))$
		\ENDFOR
		\FOR{each parameter $\theta$ in $\Theta$}
		\STATE Perform gradient descent update: $\theta \leftarrow \theta - \rho \cdot \frac{\partial \mathcal{L}}{\partial \theta}$
		\ENDFOR
		\ENDFOR
		\RETURN Optimized parameters $\Theta$
	\end{algorithmic}
\end{algorithm}

\subsection{Time complexity}
\label{sec:complexity}
Recommender systems must not only provide desirable performance in terms of accuracy and efficiency but also be optimized in terms of execution time.
Therefore, this section analyzes time complexity of the model.
\begin{itemize}
	\item 
	In the preprocessing (graph construction and denoising) step,
	time complexity of computing item similarity is 
	\( O(|\mathcal{A}| + |\mathcal{J}|^3) \), where \( |\mathcal{A}| \) is the total number of interactions and \( |\mathcal{J}| \) is the number of items. Additionally, to eliminate noise, the Louvain method is used for community detection, which has a time complexity of \( O(|\mathcal{A}| \log (|\mathcal{A}|)) \). Furthermore, the time complexity of adding new interactions based on the similarity criterion is \( O(|\mathcal{A}|) \).
	As a result, the whole time complexity of the preprocessing step is
	\( O(|\mathcal{A}| + |\mathcal{J}|^3) \).
	\item 
	In the training phase, time complexity of computing short-term embeddings is 
	\( O(L \times |\mathcal{A}| \times d) \), indicating that the computational cost of segmented graphs is equivalent to that of a complete graph. Time complexity of computing long-term embeddings, based on the sequential pattern across time intervals, is 
	\( O((T \times d^2 + T^2 \times d) \times (I + J)) \). The sequential pattern modeling based on the item sequence for computing user preference embeddings has a time complexity of 
	\( O((M \times d^2 + M^2 \times d) \times B) \), where \( B \) denotes the batch size. Finally, computing long-term embeddings using the averaging method has a time complexity of 
	\( O(T \times (I + J)) \).
	As a result, the whole time complexity of the training phase is \(
	O \Big( L \times |\mathcal{A}| \times d + (T \times d^2 + T^2 \times d) \times (I + J) + (M \times d^2 + M^2 \times d) \times B \Big)\).
\end{itemize}

\section{Experiments}
\label{sec:experiments}

In this section, we first provide an overview of the datasets, baseline models, and evaluation metrics used in our study. We then present our extensive experiments, detailing the evaluation settings and analyzing the impact of various factors on model performance.

\subsection{Experimental setup}

\subsubsection{Datasets}

We evaluate the performance of our proposed model on four datasets, which are presented in Table~\ref{tab:dataset_stats}.
\begin{itemize}
	\item \textbf{Amazon-Book}\footnote{\url{https://cseweb.ucsd.edu/~jmcauley/datasets/amazon/links.html}}: This dataset contains user ratings for Amazon books, collected in 2014 \cite{he2016ups}.
	\item \textbf{MovieLens}\footnote{\url{https://grouplens.org/datasets/movielens/10m/}}: This dataset includes user ratings for various movies within the time range of 2002 to 2009 \cite{harper2015movielens}.
	\item \textbf{Gowalla}\footnote{\url{https://snap.stanford.edu/data/loc-Gowalla.html}}: This dataset consists of users' location check-ins on the Gowalla platform in 2010 \cite{cho2011friendship}.
	\item \textbf{Yelp}\footnote{\url{https://www.yelp.com/dataset}}: This dataset contains information about businesses and user interactions with them, collected between 2009 and 2019.
\end{itemize}

\begin{table*}
	\centering
	\caption{Statistics of the datasets used in our experiments.}
	\label{tab:dataset_stats}
	\begin{tabular}{lcccc}
		\toprule
		\textbf{Dataset} & \textbf{\# users} & \textbf{\# items} & \textbf{\# interactions} & \textbf{Density} \\
		\midrule
		Amazon-book & 11199 & 30821 & 375916 & 1.1e-3 \\
		Gowalla & 48653 & 52621 & 1807125 & 7.1e-4 \\
		Movielens & 24312 & 8688 & 1758929 & 8.3e-3 \\
		Yelp & 19751 & 38391 & 1467157 & 1.9e-3 \\
		\bottomrule
	\end{tabular}
\end{table*}

\subsubsection{Baseline models}
To evaluate the performance of ALDA4Rec, we consider the following state-of-the-art recommendation models:
\begin{itemize}
	\item SASRec \cite{kang2018self}: Utilizes self-attention mechanisms to capture sequential patterns in recommender systems.
	\item TiSASRec \cite{li2020time}: Integrates time interval embeddings into a transformer-based self-attention model, dynamically adjusting attention scores based on time gaps between user interactions.
	\item LightGCN \cite{he2020lightgcn}: Simplifies graph convolution by removing non-linear transformations and feature projections. It focuses on propagating and aggregating user and item embeddings over the interaction graph, combining information from multiple propagation layers through weighted summation.
	\item GraphDA \cite{fan2023graphda}: Proposes a novel adjacency matrix integrating user-user, item-item correlations, and balanced user-item interactions via top-K sampling. It leverages pre-trained graph embeddings to enhance recommendation accuracy.
	\item DCF \cite{10.1145/3637528.3671692}: Implements a double correction strategy that enhances denoising stability. It utilizes a damping function on temporal loss values and concentration inequalities to prevent the misidentification of hard samples as noise, while employing a progressive re-labeling mechanism to maximize data exploitation.
	\item SelfGNN \cite{liu2024selfgnn}: Constructs sequential user interaction graphs and leverages GNNs to capture complex item dependencies. It also integrates self-supervised contrastive learning tasks to refine embeddings and enhance recommendation performance.
\end{itemize}

\subsubsection{Evaluation metrics}
To evaluate this method, we use two common Top-$N$ metrics: Hit Ratio $(HR)@N$ and Normalized Discounted Cumulative Gain $(NDCG)@N$. Below, we provide a brief description of these metrics. 

$NDCG@N$ is defined as follows:  

\begin{equation} 
	\text{NDCG}@N = \frac{\text{DCG}@N}{\text{IDCG}@N} \label{eq:ndcg}, 
\end{equation}  
where $DCG@N$ and $IDCG@N$ are given by:  

\begin{equation} 
	\text{DCG}@N = \sum_{i=1}^{N} \frac{2^{rel_i} - 1}{\log_2(i + 1)}. 
\end{equation}  

\begin{equation} 
	\text{IDCG}@N = \sum_{i=1}^{REL_N} \frac{2^{rel_i} - 1}{\log_2(i + 1)}. 
\end{equation}  

Here, $REL_N$ represents a stored list of the $N$ most relevant items in descending order. The function $rel_i$ denotes the relevance of the $i$-th ranked item, which is either $0$ (irrelevant) or $1$ (relevant).  

$HR@N$ measures the fraction of cases in which the ground-truth item appears among the top-$N$ recommendations. It is defined as follows:  

\begin{equation}
	\text{HR}@N = \frac{1}{|U|} \sum_{u \in U} \mathbb{I} (r_u \in R_u^N),
\end{equation}  
where $r_u$ represents the ground-truth item for user $u$, $R_u^N$ denotes the top-$N$ recommended items for user $u$, and $\mathbb{I}(\cdot)$ is an indicator function that returns $1$ if $r_u$ is in $R_u^N$ and $0$ otherwise.  

To evaluate our method, we assume $N$ is selected from $\{5, 15, 20\}$.  

\subsubsection{Hyper-parameters settings}

For the implementation of the proposed model, we utilize the TensorFlow library (version $1.14$). The model is optimized using the Adam optimizer with an initial learning rate of $1 \times 10^{-3}$, which decays by a factor of $0.96$ per epoch. The embedding dimension is set to $64$. The architecture of the GNN consists of a variable number of layers, selected from the set $\{1, 2, 3\}$. The batch size is chosen from $\{128, 256, 512\}$, ensuring efficient training. At the sequence modeling level, the number of attention layers is selected from $\{2, 3, 4\}$. Additionally, the number of short-term graphs ($T$) varies within the range $\{t \mid 3 \leq t \leq 12\}$, allowing flexibility in capturing temporal dependencies. To prevent overfitting, the $L_2$ regularization weight ($\lambda_2$) is set to $1 \times 10^{-2}$. A dropout rate of $0.5$ is applied to enhance generalization.

For the preprocessing stage, the noise-reduction weight ($\beta$), similarity threshold ($\min_{sim}$), and balancing coefficient ($\lambda_1$) are tuned via grid search on the validation set. Specifically, we explore candidate values within the interval $[0,1]$ and evaluate their combinations to identify the optimal configuration. This range is selected because these parameters act as normalized coefficients controlling noise suppression strength, similarity filtering strictness, and loss balancing, respectively. During this process, all other model parameters, including those of SelfGNN, are kept fixed to ensure a fair comparison and to isolate the impact of the preprocessing hyperparameters. The final values are selected based on the best validation performance.

\subsection{Performance comparison}
Table \ref{tab:results} compares our proposed model, ALDA4Rec, with baseline models using HR and NDCG. Among the baselines, TiSASRec \cite{li2020time} extends SASRec \cite{kang2018self} by incorporating temporal factors into its attention mechanism, while GraphDA \cite{fan2023graphda} and DCF \cite{10.1145/3637528.3671692} enhance LightGCN \cite{he2020lightgcn} through noise reduction and interaction augmentation. SelfGNN \cite{liu2024selfgnn}, a hybrid model combining GNNs and attention, achieves the best performance among baselines by learning both short- and long-term user preferences.

Our proposed model, ALDA4Rec, consistently outperforms all baseline models across all evaluation metrics. This superior performance can be attributed to its novel noise removal strategy during preprocessing and its ability to leverage the average of short-term embeddings to model long-term user relationships. Additionally, it adaptively integrates these embeddings for prediction and optimization, further refining recommendation accuracy. The "Imp" column in Table \ref{tab:results} quantifies the extent of improvement achieved by our model compared to the baselines.

Dataset-specific trends further highlight the robustness of our approach. In the Gowalla dataset, where the timeline is divided into three short-term graphs, averaging short-term embeddings results in a moderate performance improvement. In contrast, the Movielens and Amazon datasets, which feature a greater number of time intervals for short-term modeling, exhibit the most significant performance gains. These results underscore the effectiveness of our approach in leveraging temporal dynamics to enhance recommendation performance.

\begin{table*}[h!]
	\centering
	\small
	\renewcommand{\arraystretch}{0.5}		
	\caption{\centering Comparison of various recommendation models across multiple datasets. Bold values indicate the best performance, underlined values the second‑best. "Imp" shows the relative improvement of ALDA4Rec over the best baseline.}
	\vspace{5pt} 
	\begin{tabular}{lccccccc|c|c}
		
		\toprule
		\textbf{Dataset} & \textbf{Metric} & SASRec & TiSASRec & LightGCN & GraphDA & DCF & SelfGNN & ALDA4Rec & \textbf{Imp} \\
		\midrule
		\multirow{6}{*}{\textit{Movielens}} 
		& HR@5 & 0.0812  & 0.0835 & 0.1188 & 0.1243 & 0.1253 & \underline{0.1345} & \textbf{0.16} & 18.96\% \\
		& NDCG@5 & 0.0524 & 0.0531  & 0.0765 & 0.0814& 0.0903 &\underline{0.0893} & \textbf{0.106} & 18.70\% \\ \addlinespace[2pt]
		& HR@10 & 0.1361  & 0.14 & 0.1912 & 0.1994& 0.2006 & \underline{0.2103} &\textbf{0.2405}& 14.36\% \\
		& NDCG@10 & 0.07 & 0.0712  & 0.0998 & 0.1057& 0.112 & \underline{0.1139} & \textbf{0.1319} &15.80\% \\ \addlinespace[2pt]
		& HR@20 & 0.1732  & 0.1804 & 0.2886 & 0.2989& 0.3189& \underline{0.3247}& \textbf{0.3541} & 9.06\% \\
		& NDCG@20 & 0.091 & 0.0926  & 0.1243 & 0.1306& 0.1395& \underline{0.1424}& \textbf{0.1604} &12.64\% \\
		\midrule
		\multirow{6}{*}{\textit{Amazon}} 
		& HR@5  & 0.1722 & 0.1811 & 0.1672 & 0.1889& 0.1967 &\underline{0.2956} & \textbf{0.3303} & 11.74\% \\
		& NDCG@5 & 0.1421 & 0.1493 & 0.1096& 0.1371& 0.1581 &\underline{0.2108} &\textbf{0.2343} & 11.15\%  \\ \addlinespace[2pt]
		& HR@10  & 0.2678 & 0.2721 & 0.2419 & 0.2664& 0.2868 & \underline{0.3914}& \textbf{0.4339} & 10.86\% \\
		& NDCG@10 & 0.1703 & 0.1785 & 0.1336 & 0.1519& 0.1813 &\underline{0.2427} & \textbf{0.2676} & 10.26\% \\ \addlinespace[2pt]
		& HR@20  & 0.3278 & 0.3294 & 0.2724 & 0.2936& 0.3364 & \underline{0.4863} & \textbf{0.5356}& 10.14\% \\
		& NDCG@20 & 0.1853 & 0.1945 & 0.159 & 0.1714& 0.209 & \underline{0.2656} & \textbf{0.2944} & 10.84\% \\
		\midrule
		\multirow{6}{*}{\textit{Gowalla}} 
		& HR@5 & 0.4295 & 0.4412 & 0.3391 & 0.3459& 0.382& \underline{0.519} & \textbf{0.526} & 1.34\% \\
		& NDCG@5 & 0.3134 & 0.3237 & 0.2498 & 0.2573& 0.2836& \underline{0.3931} & \textbf{0.3951} & 0.50\% \\ \addlinespace[2pt]
		& HR@10 & 0.5634 & 0.5713 & 0.4690 & 0.4762& 0.501 & \underline{0.6464} & \textbf{0.6506} & 0.65\% \\
		& NDCG@10 & 0.3565 & 0.3658 & 0.2886 & 0.2963 & 0.3249& \underline{0.4345} & \textbf{0.436} & 0.34\% \\ \addlinespace[2pt]
		& HR@20 & 0.6863 & 0.6887 & 0.5981 & 0.6086 & 0.6388& \underline{0.7634} & \textbf{0.7708} & 0.96\% \\
		& NDCG@20 & 0.3909 & 0.3979 & 0.3191 & 0.3277& 0.4097& \underline{0.4632} & \textbf{0.4660} & 0.60\% \\
		\midrule
		\multirow{6}{*}{\textit{Yelp}} 
		& HR@5 & 0.0742 & 0.0745 & 0.2183 & 0.2322& 0.2322&\underline{0.2338} &  \textbf{0.2361} & 0.98\% \\
		& NDCG@5 & 0.0489 & 0.05 & 0.1416 & 0.1524& 0.1529& \underline{0.1543} & \textbf{0.1548} & 0.32\% \\ \addlinespace[2pt]
		& HR@10 & 0.104 & 0.1062 & 0.3424 & 0.3578 & \underline{0.3581}&0.3495 & \textbf{0.3638} & 1.59\% \\
		& NDCG@10 & 0.0649 & 0.0712 & 0.1816 & 0.1929& \underline{0.193}&0.1916 & \textbf{0.1956} & 1.33\% \\ \addlinespace[2pt]
		& HR@20 & 0.1541 & 0.1601 & 0.5039 & 0.5087& \underline{0.5102} &0.4964 & \textbf{0.5252} & 2.97\% \\
		& NDCG@20 & 0.0828 & 0.0831 & 0.2222 & 0.2309& \underline{0.2313} &0.2285&  \textbf{0.2366} & 2.24\% \\
		\bottomrule
	\end{tabular}
	\label{tab:results}
\end{table*}

\subsection{Ablation study}
In this section, we examine the key innovations of the proposed method, including graph construction and denoising strategies, as well as the use of short-term embedding averages to mitigate the vanishing gradient issue in GRU and address the U-shape problem in the attention mechanism.

\subsubsection{Graph construction and denoising evaluation}
\paragraph{Comparison of denoising and augmentation variants.}
\begin{table*}[!h]
	\centering
	\small
	\renewcommand{\arraystretch}{1.2}
	\caption{\centering Evaluation of graph construction and denoising on the top-10 data points. 
		Bold marks the best model and underlining the second best (consistent with Table 1).}
	\vspace{5pt}
	\begin{tabular}{lcccccccc}
		\toprule
		\multirow{2}{*}{\textbf{Model}} & 
		\multicolumn{2}{c}{\textbf{Yelp}}     & 
		\multicolumn{2}{c}{\textbf{Movielens}}& 
		\multicolumn{2}{c}{\textbf{Amazon}}   & 
		\multicolumn{2}{c}{\textbf{Gowalla}}  \\
		\cmidrule(lr){2-3}\cmidrule(lr){4-5}\cmidrule(lr){6-7}\cmidrule(lr){8-9}
		& HR  & NDCG & HR  & NDCG & HR  & NDCG & HR  & NDCG \\
		\midrule
		Noisy-data & 0.1855 & 0.3475 & 0.1088 & 0.2088 & 0.2189 & 0.3615 & 0.4349 & 0.6453 \\
		GraphDA    & 0.1929 & 0.3578 & 0.1057 & 0.1994 & 0.1519 & 0.2664 & 0.2963 & 0.4762 \\
		DCF   & 0.193 & 0.3581 & 0.112 & 0.2006 & 0.1813 & 0.2868 & 0.3249 & 0.501 \\
		SelfGNN    & 0.1916 & 0.3495 & 0.1139 & 0.2103 & 0.2427 & 0.3914 & 0.4345 & 0.6464 \\
		\midrule
		DA4Rec     & 0.1921 & 0.3591 &
		0.1218 & 0.2234 &
		0.2580 & 0.4154 &
		0.4360 & 0.6503 \\
		D4Rec  & 0.1913 & 0.3480 &
		0.1216 & 0.2231 &
		0.2427 & 0.4057 &
		0.4352 & 0.6459 \\
		A4Rec  & 0.1916 & 0.3512 &
		0.1091 & 0.2104 &
		0.2206 & 0.3881 &
		0.4360 & 0.6498 \\
		\bottomrule
	\end{tabular}
	\label{table:comparison-test2}
\end{table*}

To evaluate the performance of our community-based denoising method, we compare the following approaches: (i) a noisy‑data model, which is a variant of SelfGNN without self‑supervised noise removal; (ii) DA4Rec\footnote{DA4Rec stands for Denoising and Augmentation for Recommendation}, a version of our model that uses only the preprocessing step for graph construction and denoising, thereby replacing SelfGNN’s self‑supervised loss with our preprocessing strategy; (iii) the original SelfGNN model; (iv) GraphDA; (v) DCF; (vi) D4Rec\footnote{D4Rec stands for Denoising for Recommendation}, which applies only the denoising component of our preprocessing; and (vii) A4Rec\footnote{A4Rec stands for Augmentation for Recommendation}, which applies only the augmentation component. To decouple the effects of denoising and augmentation explicitly, D4Rec merely prunes noisy interactions, whereas A4Rec only augments the graph by adding similar items.

This design allows us to analyze the individual contribution of each component. As shown in Table~\ref{table:comparison-test2}, the two components affect datasets differently: on Movielens, D4Rec achieves most of the performance gain, indicating that denoising dominates; on Gowalla, however, A4Rec contributes more significantly, showing that augmentation is particularly beneficial in sparse settings. Overall, both denoising and augmentation positively influence performance, and their combination in DA4Rec leads to the best results.
Moreover, the results show that DA4Rec outperforms all other methods. Not only does DA4Rec surpass SelfGNN in noise removal, but it also reduces complexity by addressing noise during the preprocessing stage. 

\paragraph{Robustness to noisy interactions.}
\begin{figure*} [!h]
	\centering
	\subfloat[Amazon]
	{
		\includegraphics[scale=.38]{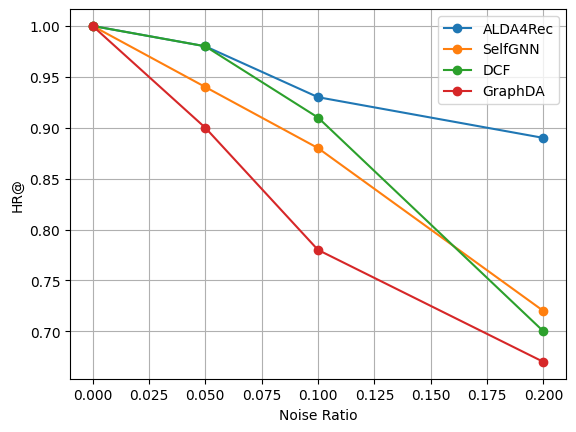}
		
	}
	\subfloat[Movielens]
	{
		\includegraphics[scale=.38]{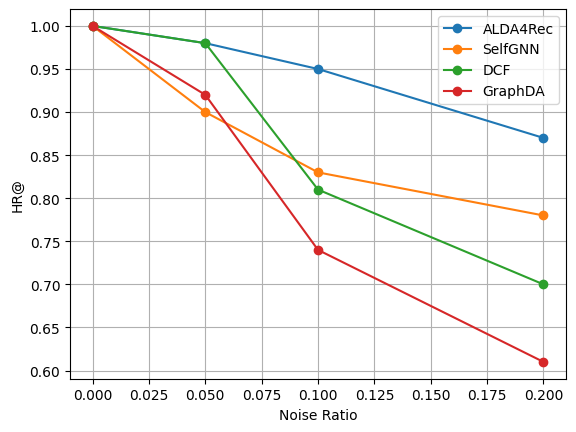}
	}
	\caption{Performance comparison under varying noise ratios in terms of HR@10 on the Amazon and MovieLens datasets.\label{fig:nr}}
\end{figure*}

	To evaluate the robustness of ALDA4Rec under noisy conditions, we conduct experiments by injecting synthetic noise into the interaction sequences. Specifically, we randomly replace a certain proportion of real user–item interactions with randomly generated fake items for each user, and then retrain the model on these corrupted sequences. We consider noise ratios of 5\%, 10\%, and 20\% to simulate varying levels of data corruption.	For comparison, we select SelfGNN, GraphDA, and DCF as baseline models, as they incorporate effective noise-handling or detection mechanisms. As illustrated in Figure \ref{fig:nr}, all models exhibit performance degradation as the noise level increases; however, ALDA4Rec consistently demonstrates superior robustness across all settings. We attribute this advantage to ALDA4Rec’s ability to mitigate noise during the preprocessing stage, where it effectively filters or attenuates unreliable interactions before model training. This early-stage noise handling enables the model to maintain more stable representations and reduces the propagation of corrupted signals, resulting in improved resilience under noisy conditions.
\paragraph{Evaluation of the Proposed Similarity Measure.}
\begin{table*}[!t]
	\centering
	\small
	\renewcommand{\arraystretch}{1.0}
	\setlength{\tabcolsep}{5pt}
	\caption{\centering Performance comparison of different similarity matrices on the Movielens dataset for top-10 recommendation. We report HR@10 and NDCG@10 on the full item set, as well as on head/tail item partitions and cold user groups. Items are ranked by interaction frequency, where the bottom 20\% are defined as tail items and the remaining 80\% as head items. Users with fewer than five interactions are categorized as cold users (1,186 out of 24,312).}
	\vspace{5pt}
	\begin{tabular}{lrrrrrrrr}
		\toprule
		\makecell{\textbf{Similarity} \\ \textbf{Matrix}} & 
		\textbf{HR} & \textbf{NDCG} & \
		\makecell{\textbf{HR} \\ \textbf{(Tail)}} & \makecell{\textbf{NDCG} \\ \textbf{(Tail)}} &
		\makecell{\textbf{HR} \\ \textbf{(Head)}} & \makecell{\textbf{NDCG} \\ \textbf{(Head)}} &
		\makecell{\textbf{HR} \\ \textbf{(Cold)}} & \makecell{\textbf{NDCG} \\ \textbf{(Cold)}} \\
		\midrule
		Cosine    & 0.2097 & 0.1106 & 0.0000 & 0.0000 & 0.2097 & 0.1106  & 0.1577 &  0.0846\\
		Jaccard   & 0.2103 & 0.1215 & 0.0000 & 0.0000 & 0.2103 & 0.1215  & 0.1772 & 0.0889 \\
		Ours      & 0.2405 & 0.1319 & 0.0122 & 0.0076 & 0.2303 & 0.1243  & 0.1652 &  0.0876 \\
		\bottomrule
	\end{tabular}
	\label{tab:sim_table}
\end{table*}

To empirically examine whether the proposed similarity formulation in Eq.~\eqref{eq:similarity} mitigates popularity bias, we conducted an additional analysis on the head and tail item subsets of the Movielens dataset. Items were ranked according to their interaction frequency, where the bottom 20\% were categorized as \emph{tail} items and the remaining 80\% as \emph{head} items. This partition results in 1,738 tail items and 8,688 head items. To isolate the effect of the similarity function, we replaced only the similarity component while keeping all other parts of the recommendation framework unchanged. The proposed similarity metric was compared against two widely used baselines, namely Cosine and Jaccard similarity. We report HR@10 and NDCG@10 on the full item set, as well as separately on the head and tail subsets.

As shown in Table~\ref{tab:sim_table}, the proposed method achieves the best overall recommendation performance on the full item set. More importantly, while both Cosine and Jaccard similarities yield zero performance on the tail subset, our method attains non-zero HR@10 and NDCG@10 values. This result indicates that the proposed similarity formulation is more effective at capturing meaningful relationships among infrequent and less popular items. These findings suggest that the proposed similarity measure alleviates popularity bias without sacrificing recommendation quality on frequently interacted items. In particular, it improves exposure to tail items while maintaining competitive performance on head items.

We also evaluated performance for \emph{cold users}, defined as users with fewer than five historical interactions. In this setting, Jaccard similarity achieves the strongest performance, whereas the proposed method ranks second and still outperforms Cosine similarity. This observation suggests that although the proposed similarity is particularly beneficial for mitigating item popularity bias, highly sparse user histories may favor overlap-based similarity measures such as Jaccard.
\paragraph{Preprocessing cost and scalability.}
\begin{table*}[!h]
	\centering
	\small
	\renewcommand{\arraystretch}{1.0}   
	\setlength{\tabcolsep}{4pt}         
	\caption{\centering
		Specifications of the preprocessing step across all datasets.
		"Execution time" is the total time for computing similarity scores,
		community detection, identifying noisy edges and augmenting new edges.}
	\vspace{5pt}
	\begin{tabular}{lrrrr}
		\toprule
		\textbf{Dataset} &
		\makecell{\textbf{\#initial}\\\textbf{interactions}} &
		\makecell{\textbf{\#noisy}\\\textbf{interactions}} &
		\makecell{\textbf{\#augmented}\\\textbf{interactions}} &
		\makecell{\textbf{Execution}\\\textbf{time (min)}} \\
		\midrule
		Movielens & 1{,}758{,}929 & 295{,}493 & 378    & 36 \\
		Amazon    &   375{,}916   & 176{,}311 & 43{,}562 & 13 \\
		Yelp      & 1{,}467{,}157 & 487{,}987 & 8{,}570 & 50 \\
		Gowalla   & 1{,}807{,}125 & 389{,}540 & 89     & 48 \\
		\bottomrule
	\end{tabular}
	\label{tab:data_table}
\end{table*}

Table \ref{tab:data_table} presents the details of the preprocessing step, including noise removal and the addition of new interactions. One challenge of our proposed model is the computational time required for preprocessing, which increases with the number of interactions and items.
In particular and as discussed in Section \ref{sec:complexity}, computing similarity scores has a quadratic time complexity with respect to the number of items. However, as shown in Table \ref{tab:data_table}, the entire preprocessing phase can be completed in less than an hour for all datasets.
It is also important to emphasize that this computational overhead is confined to the preprocessing phase and does not directly impact the system's performance during subsequent stages.
\subsubsection{Long-term embedding evaluation via averaging}
\begin{table*}[!h]
	\centering
	\small
	\renewcommand{\arraystretch}{1.2}
	\setlength{\tabcolsep}{5pt}
	\caption{\centering
		Evaluation of long-term embedding addition (simple averaging) on the top-10
		recommendations.  Bold marks the best score; underlining the second-best.}
	\vspace{5pt}
	\begin{tabular}{lcccccccc}
		\toprule
		\multirow{2}{*}{\textbf{Model}} &
		\multicolumn{2}{c}{\textbf{Yelp}} &
		\multicolumn{2}{c}{\textbf{Movielens}} &
		\multicolumn{2}{c}{\textbf{Amazon}} &
		\multicolumn{2}{c}{\textbf{Gowalla}} \\
		\cmidrule(lr){2-3}\cmidrule(lr){4-5}\cmidrule(lr){6-7}\cmidrule(lr){8-9}
		& HR & NDCG & HR & NDCG & HR & NDCG & HR & NDCG \\
		\midrule
		AL4Rec   & {0.1920} & {0.3548} &
		{0.1257} & {0.2305} &
		0.2343 & {0.4029} &
		0.4340 & 0.6454 \\
		SelfGNN  & 0.1916 & 0.3495 &
		0.1139 & 0.2103 &
		{0.2427} & 0.3914 &
		{0.4345} & {0.6464} \\
		AL-SGL   & {0.2031} & {0.3743} &
		{0.1273} &{0.2354} &
		{0.2582} & {0.4120} &
		{0.4356} & {0.6480} \\
		\bottomrule
	\end{tabular}
	\label{table:comparison-test1}
\end{table*}

To evaluate the impact of aggregating long-term embeddings via averaging, we compare the following methods in this section: i) AL4Rec,\footnote{AL4Rec is an abbreviation for Adaptive Long-term Embedding for Recommendation} which represents our model without the preprocessing phase (i.e., noise reduction and augmentation), ii) AL-SGL,\footnote{AL-SGL is an abbreviation for Adaptive Long-term Embedding and SelfGNN Loss} which incorporates our proposed mean-level long-term embedding aggregation into the SelfGNN model, and iii) the SelfGNN model.
The results are presented in Table \ref{table:comparison-test1}. AL-SGL outperforms the other methods across all datasets, reflecting the impact of using the mean function for aggregating embeddings.

By analyzing the results presented in Tables \ref{tab:results},  \ref{table:comparison-test2} and \ref{table:comparison-test1}, it is evident that the AL-SGL model performs better for the Yelp dataset. This superior performance can be attributed to the division of the dataset into 12 time intervals to model short-term interactions. Incorporating the average of short-term embeddings ensures that the model captures information from all intervals effectively.
Additionally, given that the average number of user interactions in most intervals is approximately 4, the community detection method struggles to effectively remove noisy data. In contrast, AL-SGL, by integrating self-supervised learning (from SelfGNN) with adaptive long-term embeddings, demonstrates superior noise reduction capabilities.

Furthermore, a comparison of our proposed model's preprocessing method (AL4Rec) with SelfGNN for the Yelp dataset indicates that AL4Rec performs better. This improvement arises from the inclusion of average embeddings. Without these, the initial and final embeddings disproportionately influence predictions due to the U-shape issue. Notably, the highest number of interactions occurs in the final three short-term intervals, where each user has an average of 10 interactions in the final embeddings. Under such conditions, our noise removal method outperforms self-supervised learning.

\subsection{Hyper-parameter analysis}

\begin{figure*}[!h]
	\centering
	\subfloat[Amazon]
	{
		\includegraphics[scale=.38]{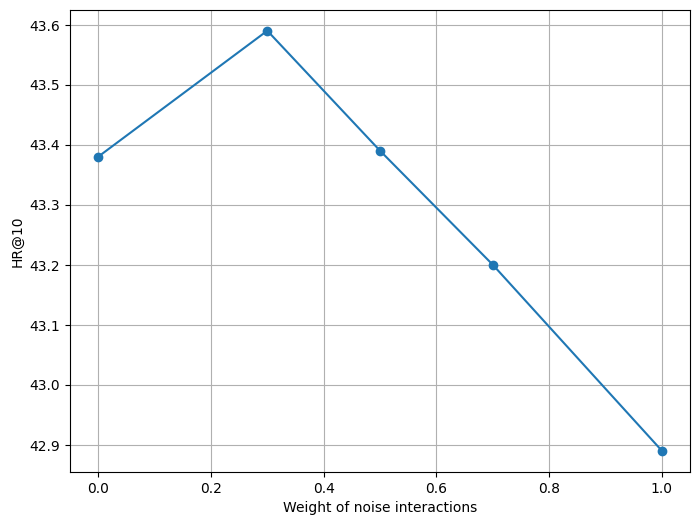}
		
	}
	\subfloat[Movielens]
	{
		\includegraphics[scale=.38]{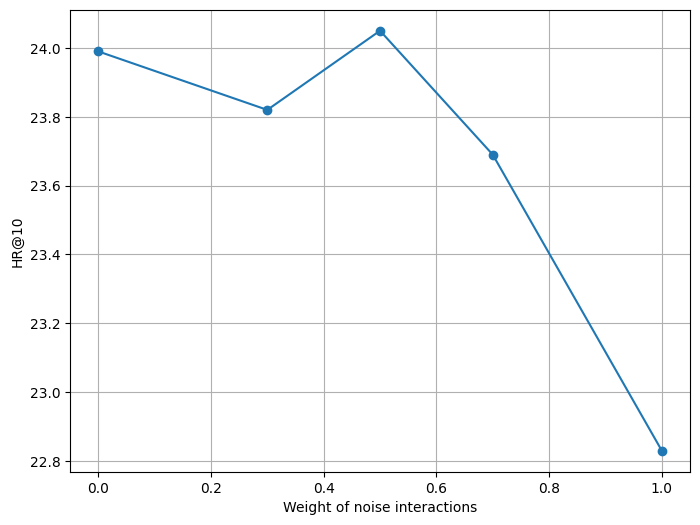}
	}
	\caption{Analysis of the Performance of Our Model for Various Values of $\beta$ in HR@10.\label{fig:boost}}	
\end{figure*}

This section examines the impact of key hyper-parameters on model performance, specifically perturbation weight ($\beta$), similarity threshold ($min_{sim}$), and $\lambda_1$. Figure \ref{fig:boost} illustrates that the optimal $\beta$ values are 0.5 for Movielens and 0.3 for Amazon. These results indicate that our model performs the worst when noise detection is disabled ($\beta = 0$), as it fails to mitigate the influence of noisy interactions in user behavior modeling. However, completely discarding detected noisy interactions is also detrimental; instead, they should be considered with a reduced weight to better reflect user behavior.

\begin{figure*} [!h]
	\centering
	\subfloat[Amazon]
	{
		\includegraphics[scale=.38]{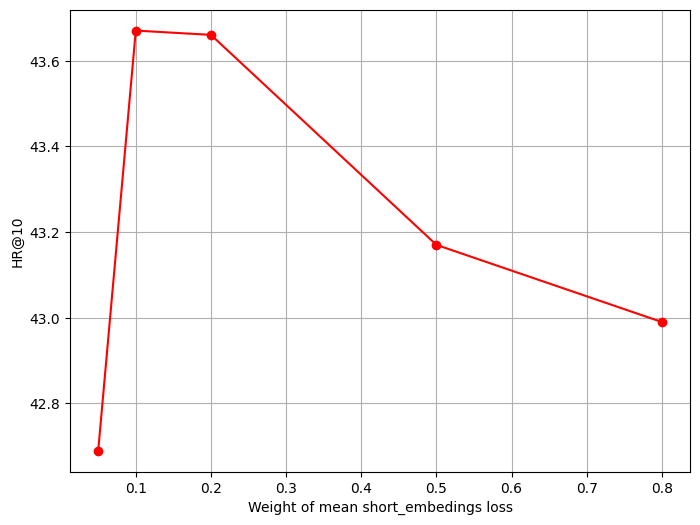}
		
	}
	\subfloat[Movielens]
	{
		\includegraphics[scale=.38]{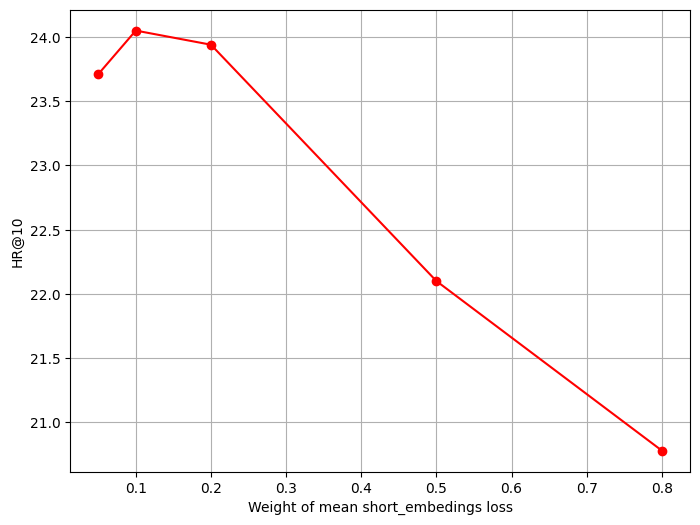}
	}
	\caption{Performance of our model for various values of $\lambda_1$ in HR@10.\label{fig:omg}}	
\end{figure*}

\begin{figure*}[!h]
	\centering
	\subfloat[Amazon]
	{
		\includegraphics[scale=.38]{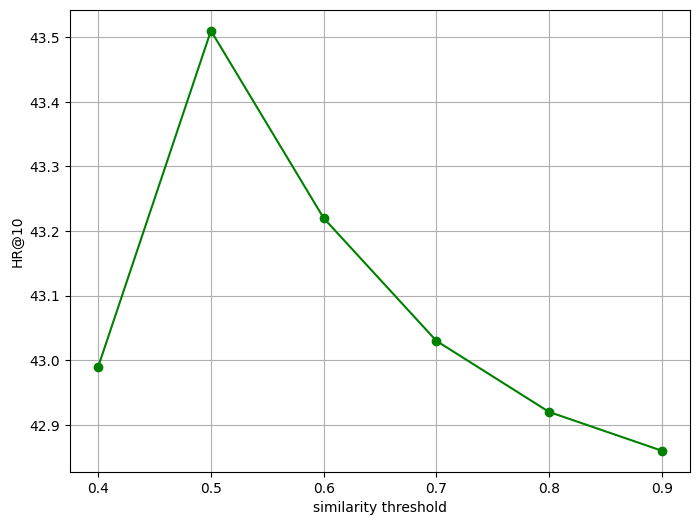}
		
	}
	\subfloat[Movielens]
	{
		\includegraphics[scale=.38]{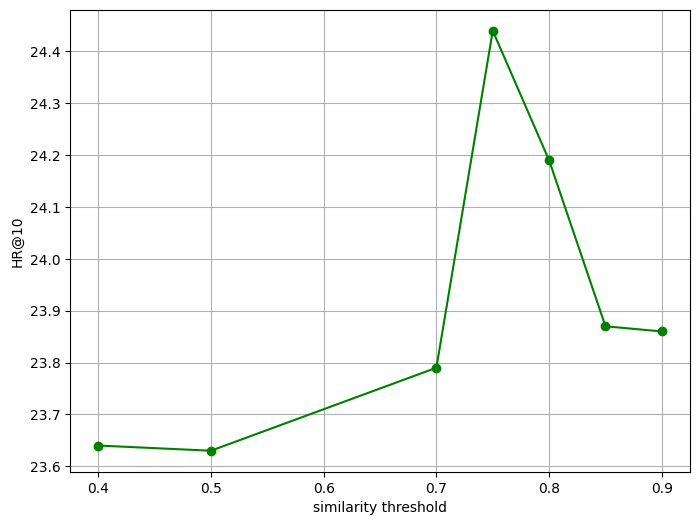}
	}
	\caption{Impact of $min_{sim}$ variation on model performance in HR@10.\label{fig:min_sim}}	
\end{figure*}

Figure \ref{fig:omg} shows that the optimal $\lambda_1$ is 0.1, suggesting that placing greater emphasis on optimizing the mean-level long-term embedding enhances model performance. Figure \ref{fig:min_sim} highlights that the best values for $min_{sim}$ are 0.75 for Movielens and 0.5 for Amazon. Models trained on datasets with fewer interactions and higher sparsity benefit from a lower $min_{sim}$, as it allows more interactions to be incorporated, leading to better user behavior modeling. These findings underscore the importance of carefully tuning hyper-parameters to achieve optimal model performance.

\section{Conclusion}
\label{sec:conclusion}

In this paper, we proposed a sequential recommendation system leveraging GNNs to enhance recommendation accuracy. Our approach offers two primary contributions. First, we introduced a novel graph construction and denoising technique that refines user interaction data by reducing noise and effectively adding interactions with overlooked items that closely match user preferences. Second, we developed a model designed to overcome the vanishing gradient problem in GRU and address the U-shaped distribution issue observed in attention mechanisms. By integrating short-term embeddings through GRUs, attention layers, and averaging methods to construct long-term representations, our approach effectively preserved historical user interactions, leading to a substantial improvement in prediction accuracy. 

Future research includes: (i) enhancing the system's adaptability by directly incorporating immediate short‑term user preferences into the prediction process, allowing for greater responsiveness to dynamically evolving user interests; and (ii) supplementing community detection with other network analysis techniques, such as centrality algorithms \cite{DBLP:conf/cikm/ChehreghaniBA19}, to further improve noise detection.


\bibliographystyle{plain}

\bibliography{references} 

@ARTICLE{pbiloss,
      title={{PBiLoss}: Popularity-Aware Regularization to Improve Fairness in Graph-Based Recommender Systems}, 
      author={Mohammad Naeimi and Mostafa Haghir Chehreghani},
      year={2026},
      journal = {Appl. Soft Comput.}, 
      volume={},
      pages={},
      note = {to appear}	
 }

@inproceedings{DBLP:conf/cikm/ChehreghaniBA19,
  author       = {Mostafa Haghir Chehreghani and
                  Albert Bifet and
                  Talel Abdessalem},
  editor       = {Wenwu Zhu and
                  Dacheng Tao and
                  Xueqi Cheng and
                  Peng Cui and
                  Elke A. Rundensteiner and
                  David Carmel and
                  Qi He and
                  Jeffrey Xu Yu},
  title        = {Adaptive Algorithms for Estimating Betweenness and \emph{k}-path Centralities},
  booktitle    = {Proceedings of the 28th {ACM} International Conference on Information
                  and Knowledge Management, {CIKM} 2019, Beijing, China, November 3-7,
                  2019},
  pages        = {1231--1240},
  year         = {2019},
  url          = {https://doi.org/10.1145/3357384.3358064},
  doi          = {10.1145/3357384.3358064},
  bibsource    = {dblp computer science bibliography, https://dblp.org}
}

@ARTICLE{maskdiffusion,
  author={Li, Kaibei and Zhang, Yihao and Li, Xiaokang and Yuan, Meng and Zhou, Wei},
  journal={IEEE Transactions on Knowledge and Data Engineering}, 
  title={Mask Diffusion-Based Contrastive Learning for Knowledge-Aware Recommendation}, 
  year={2025},
  volume={37},
  number={9},
  pages={5407-5419},
  doi={10.1109/TKDE.2025.3582767}
}

@article{prototypelearning,
title = {Prototype learning based hierarchical decoupling for multimodal recommendation},
journal = {Expert Systems with Applications},
volume = {304},
pages = {130763},
year = {2026},
issn = {0957-4174},
doi = {https://doi.org/10.1016/j.eswa.2025.130763},
url = {https://www.sciencedirect.com/science/article/pii/S0957417425043787},
author = {Jiangchuan Liu and Yihao Zhang and Qinyang He and Ran Yang and Xibin Wang and Wei Zhou}
}

@ARTICLE{latentdiffusion,
  author={He, Qinyang and Zhang, Yihao and Li, Kaibei and Li, Xiaokang and Zhou, Wei},
  journal={IEEE Transactions on Systems, Man, and Cybernetics: Systems}, 
  title={Latent Diffusion Model for Social Recommendation}, 
  year={2026},
  volume={56},
  number={5},
  pages={3355-3369},
  doi={10.1109/TSMC.2026.3657816}
}

@article{smoothdiffusion,
title = {Smooth diffusion model for multimodal recommendation},
journal = {Knowledge-Based Systems},
volume = {331},
pages = {114807},
year = {2026},
issn = {0950-7051},
doi = {https://doi.org/10.1016/j.knosys.2025.114807},
url = {https://www.sciencedirect.com/science/article/pii/S0950705125018453},
author = {Qinyang He and Kaibei Li and Yihao Zhang and Xiaokang Li and Wei Zhou}
}

@article{hierarchydiffusion,
author = {Li, Kaibei and Zhang, Yihao and He, Qinyang and Li, Xiaokang},
year = {2026},
month = {01},
pages = {},
title = {Leveraging hierarchy-aware diffusion model and knowledge-enhanced contrastive learning for recommendation},
volume = {68},
journal = {Knowledge and Information Systems},
doi = {10.1007/s10115-025-02639-4}
}

@article{barbero2024transformers,
	author       = {Federico Barbero and
	Andrea Banino and
	Steven Kapturowski and
	Dharshan Kumaran and
	Jo{\~{a}}o G. M. Ara{\'{u}}jo and
	Alex Vitvitskyi and
	Razvan Pascanu and
	Petar Velickovic},
	title        = {Transformers need glasses! Information over-squashing in language
	tasks},
	journal      = {CoRR},
	volume       = {abs/2406.04267},
	year         = {2024},
	url          = {https://doi.org/10.48550/arXiv.2406.04267},
	doi          = {10.48550/ARXIV.2406.04267},
	eprinttype    = {arXiv},
	eprint       = {2406.04267},
	bibsource    = {dblp computer science bibliography, https://dblp.org}
}

@misc{gholinejad2025disentanglingpopularityqualityedge,
      title={Disentangling Popularity and Quality: An Edge Classification Approach for Fair Recommendation}, 
      author={Nemat Gholinejad and Mostafa Haghir Chehreghani},
      year={2025},
      eprint={2502.15699},
      archivePrefix={arXiv},
      primaryClass={cs.IR},
      url={https://arxiv.org/abs/2502.15699}, 
}

@inproceedings{DBLP:conf/iclr/KipfW17,
  author       = {Thomas N. Kipf and
                  Max Welling},
  title        = {Semi-Supervised Classification with Graph Convolutional Networks},
  booktitle    = {5th International Conference on Learning Representations, {ICLR} 2017,
                  Toulon, France, April 24-26, 2017, Conference Track Proceedings},
  year         = {2017},
  url          = {https://openreview.net/forum?id=SJU4ayYgl},
  bibsource    = {dblp computer science bibliography, https://dblp.org}
}

@article{DBLP:journals/natmi/Chehreghani22,
  author       = {Mostafa Haghir Chehreghani},
  title        = {Half a decade of graph convolutional networks},
  journal      = {Nat. Mach. Intell.},
  volume       = {4},
  number       = {3},
  pages        = {192--193},
  year         = {2022},
  url          = {https://doi.org/10.1038/s42256-022-00466-8},
  doi          = {10.1038/S42256-022-00466-8},
  bibsource    = {dblp computer science bibliography, https://dblp.org}
}

@article{10.1145/3700790,
author = {Gholamzadeh Nasrabadi, Fatemeh and Kashani, Amirhossein and Zahedi, Pegah and Haghir Chehreghani, Mostafa},
title = {Content Augmented Graph Neural Networks},
year = {2024},
publisher = {Association for Computing Machinery},
address = {New York, NY, USA},
issn = {1559-1131},
url = {https://doi.org/10.1145/3700790},
doi = {10.1145/3700790},
journal = {ACM Trans. Web},
month = oct
}

@article{DBLP:journals/tjs/ZohrabiSC24,
  author       = {Mohammadjavad Zohrabi and
                  Saeed Saravani and
                  Mostafa Haghir Chehreghani},
  title        = {Centrality-based and similarity-based neighborhood extension in graph
                  neural networks},
  journal      = {J. Supercomput.},
  volume       = {80},
  number       = {16},
  pages        = {24638--24663},
  year         = {2024},
  url          = {https://doi.org/10.1007/s11227-024-06336-x},
  doi          = {10.1007/S11227-024-06336-X},
  bibsource    = {dblp computer science bibliography, https://dblp.org}
}

@article{DBLP:journals/corr/abs-2411-15671,
	author       = {Ali Behrouz and
	Ali Parviz and
	Mahdi Karami and
	Clayton Sanford and
	Bryan Perozzi and
	Vahab Mirrokni},
	title        = {Best of Both Worlds: Advantages of Hybrid Graph Sequence Models},
	journal      = {CoRR},
	volume       = {abs/2411.15671},
	year         = {2024},
	url          = {https://doi.org/10.48550/arXiv.2411.15671},
	doi          = {10.48550/ARXIV.2411.15671},
	eprinttype    = {arXiv},
	eprint       = {2411.15671},
	bibsource    = {dblp computer science bibliography, https://dblp.org}
}

@inproceedings{he2016ups,
	author       = {Ruining He and
	Julian J. McAuley},
	editor       = {Jacqueline Bourdeau and
	Jim Hendler and
	Roger Nkambou and
	Ian Horrocks and
	Ben Y. Zhao},
	title        = {Ups and Downs: Modeling the Visual Evolution of Fashion Trends with
	One-Class Collaborative Filtering},
	booktitle    = {Proceedings of the 25th International Conference on World Wide Web,
	{WWW} 2016, Montreal, Canada, April 11 - 15, 2016},
	pages        = {507--517},
	year         = {2016},
	url          = {https://doi.org/10.1145/2872427.2883037},
	doi          = {10.1145/2872427.2883037},
	bibsource    = {dblp computer science bibliography, https://dblp.org}
}

@article{harper2015movielens,
	author       = {F. Maxwell Harper and
	Joseph A. Konstan},
	title        = {The MovieLens Datasets: History and Context},
	journal      = {{ACM} Trans. Interact. Intell. Syst.},
	volume       = {5},
	number       = {4},
	pages        = {19:1--19:19},
	year         = {2016},
	url          = {https://doi.org/10.1145/2827872},
	doi          = {10.1145/2827872},
	bibsource    = {dblp computer science bibliography, https://dblp.org}
}

@inproceedings{cho2011friendship,
	author       = {Eunjoon Cho and
	Seth A. Myers and
	Jure Leskovec},
	editor       = {Chid Apt{\'{e}} and
	Joydeep Ghosh and
	Padhraic Smyth},
	title        = {Friendship and mobility: user movement in location-based social networks},
	booktitle    = {Proceedings of the 17th {ACM} {SIGKDD} International Conference on
	Knowledge Discovery and Data Mining, San Diego, CA, USA, August 21-24,
	2011},
	pages        = {1082--1090},
	year         = {2011},
	url          = {https://doi.org/10.1145/2020408.2020579},
	doi          = {10.1145/2020408.2020579},
	bibsource    = {dblp computer science bibliography, https://dblp.org}
}

@inproceedings{kang2018self,
	author       = {Wang{-}Cheng Kang and
	Julian J. McAuley},
	title        = {Self-Attentive Sequential Recommendation},
	booktitle    = {{IEEE} International Conference on Data Mining, {ICDM} 2018, Singapore,
	November 17-20, 2018},
	pages        = {197--206},
	year         = {2018},
	url          = {https://doi.org/10.1109/ICDM.2018.00035},
	doi          = {10.1109/ICDM.2018.00035},
	bibsource    = {dblp computer science bibliography, https://dblp.org}
}

@inproceedings{li2020time,
	author       = {Jiacheng Li and
	Yujie Wang and
	Julian J. McAuley},
	editor       = {James Caverlee and
	Xia (Ben) Hu and
	Mounia Lalmas and
	Wei Wang},
	title        = {Time Interval Aware Self-Attention for Sequential Recommendation},
	booktitle    = {{WSDM} '20: The Thirteenth {ACM} International Conference on Web Search
	and Data Mining, Houston, TX, USA, February 3-7, 2020},
	pages        = {322--330},
	year         = {2020},
	url          = {https://doi.org/10.1145/3336191.3371786},
	doi          = {10.1145/3336191.3371786},
	bibsource    = {dblp computer science bibliography, https://dblp.org}
}

@inproceedings{he2020lightgcn,
	author       = {Xiangnan He and
	Kuan Deng and
	Xiang Wang and
	Yan Li and
	Yong{-}Dong Zhang and
	Meng Wang},
	editor       = {Jimmy X. Huang and
	Yi Chang and
	Xueqi Cheng and
	Jaap Kamps and
	Vanessa Murdock and
	Ji{-}Rong Wen and
	Yiqun Liu},
	title        = {LightGCN: Simplifying and Powering Graph Convolution Network for Recommendation},
	booktitle    = {Proceedings of the 43rd International {ACM} {SIGIR} conference on
	research and development in Information Retrieval, {SIGIR} 2020, Virtual
	Event, China, July 25-30, 2020},
	pages        = {639--648},
	year         = {2020},
	url          = {https://doi.org/10.1145/3397271.3401063},
	doi          = {10.1145/3397271.3401063},
	bibsource    = {dblp computer science bibliography, https://dblp.org}
}

@inproceedings{fan2023graphda,
	author       = {Ziwei Fan and
	Ke Xu and
	Zhang Dong and
	Hao Peng and
	Jiawei Zhang and
	Philip S. Yu},
	editor       = {Hsin{-}Hsi Chen and
	Wei{-}Jou (Edward) Duh and
	Hen{-}Hsen Huang and
	Makoto P. Kato and
	Josiane Mothe and
	Barbara Poblete},
	title        = {Graph Collaborative Signals Denoising and Augmentation for Recommendation},
	booktitle    = {Proceedings of the 46th International {ACM} {SIGIR} Conference on
	Research and Development in Information Retrieval, {SIGIR} 2023, Taipei,
	Taiwan, July 23-27, 2023},
	pages        = {2037--2041},
	year         = {2023},
	url          = {https://doi.org/10.1145/3539618.3591994},
	doi          = {10.1145/3539618.3591994},
	bibsource    = {dblp computer science bibliography, https://dblp.org}
}

@inproceedings{liu2024selfgnn,
	author = {Liu, Yuxi and Xia, Lianghao and Huang, Chao},
	title = {SelfGNN: Self-Supervised Graph Neural Networks for Sequential Recommendation},
	year = {2024},
	isbn = {9798400704314},
	publisher = {Association for Computing Machinery},
	address = {New York, NY, USA},
	url = {https://doi.org/10.1145/3626772.3657716},
	doi = {10.1145/3626772.3657716},
	booktitle = {Proceedings of the 47th International ACM SIGIR Conference on Research and Development in Information Retrieval},
	pages = {1609–1618},
	numpages = {10},
	location = {Washington DC, USA},
	series = {SIGIR '24}
}

@article{louvain_community,
	author       = {Vincent D. Blondel and
	Jean{-}Loup Guillaume and
	Renaud Lambiotte},
	title        = {Fast unfolding of communities in large networks: 15 years later},
	journal      = {CoRR},
	volume       = {abs/2311.06047},
	year         = {2023},
	url          = {https://doi.org/10.48550/arXiv.2311.06047},
	doi          = {10.48550/ARXIV.2311.06047},
	eprinttype    = {arXiv},
	eprint       = {2311.06047},
	bibsource    = {dblp computer science bibliography, https://dblp.org}
}

@article{GC-MC,
	author       = {Rianne van den Berg and
	Thomas N. Kipf and
	Max Welling},
	title        = {Graph Convolutional Matrix Completion},
	journal      = {CoRR},
	volume       = {abs/1706.02263},
	year         = {2017},
	url          = {http://arxiv.org/abs/1706.02263},
	eprinttype    = {arXiv},
	eprint       = {1706.02263},
	bibsource    = {dblp computer science bibliography, https://dblp.org}
}

@inproceedings{NGCF,
	author       = {Xiang Wang and
	Xiangnan He and
	Meng Wang and
	Fuli Feng and
	Tat{-}Seng Chua},
	editor       = {Benjamin Piwowarski and
	Max Chevalier and
	{\'{E}}ric Gaussier and
	Yoelle Maarek and
	Jian{-}Yun Nie and
	Falk Scholer},
	title        = {Neural Graph Collaborative Filtering},
	booktitle    = {Proceedings of the 42nd International {ACM} {SIGIR} Conference on
	Research and Development in Information Retrieval, {SIGIR} 2019, Paris,
	France, July 21-25, 2019},
	pages        = {165--174},
	year         = {2019},
	url          = {https://doi.org/10.1145/3331184.3331267},
	doi          = {10.1145/3331184.3331267},
	bibsource    = {dblp computer science bibliography, https://dblp.org}
}

@inproceedings{GCCF,
	author       = {Lei Chen and
	Le Wu and
	Richang Hong and
	Kun Zhang and
	Meng Wang},
	title        = {Revisiting Graph Based Collaborative Filtering: {A} Linear Residual
	Graph Convolutional Network Approach},
	booktitle    = {The Thirty-Fourth {AAAI} Conference on Artificial Intelligence, {AAAI}
	2020, The Thirty-Second Innovative Applications of Artificial Intelligence
	Conference, {IAAI} 2020, The Tenth {AAAI} Symposium on Educational
	Advances in Artificial Intelligence, {EAAI} 2020, New York, NY, USA,
	February 7-12, 2020},
	pages        = {27--34},
	year         = {2020},
	url          = {https://doi.org/10.1609/aaai.v34i01.5330},
	doi          = {10.1609/AAAI.V34I01.5330},
	bibsource    = {dblp computer science bibliography, https://dblp.org}
}

@inproceedings{UltraGCN,
	author       = {Kelong Mao and
	Jieming Zhu and
	Xi Xiao and
	Biao Lu and
	Zhaowei Wang and
	Xiuqiang He},
	editor       = {Gianluca Demartini and
	Guido Zuccon and
	J. Shane Culpepper and
	Zi Huang and
	Hanghang Tong},
	title        = {UltraGCN: Ultra Simplification of Graph Convolutional Networks for
	Recommendation},
	booktitle    = {{CIKM} '21: The 30th {ACM} International Conference on Information
	and Knowledge Management, Virtual Event, Queensland, Australia, November
	1 - 5, 2021},
	pages        = {1253--1262},
	year         = {2021},
	url          = {https://doi.org/10.1145/3459637.3482291},
	doi          = {10.1145/3459637.3482291},
	bibsource    = {dblp computer science bibliography, https://dblp.org}
}

@inproceedings{SR-GNN,
	author       = {Shu Wu and
	Yuyuan Tang and
	Yanqiao Zhu and
	Liang Wang and
	Xing Xie and
	Tieniu Tan},
	title        = {Session-Based Recommendation with Graph Neural Networks},
	booktitle    = {The Thirty-Third {AAAI} Conference on Artificial Intelligence, {AAAI}
	2019, The Thirty-First Innovative Applications of Artificial Intelligence
	Conference, {IAAI} 2019, The Ninth {AAAI} Symposium on Educational
	Advances in Artificial Intelligence, {EAAI} 2019, Honolulu, Hawaii,
	USA, January 27 - February 1, 2019},
	pages        = {346--353},
	year         = {2019},
	url          = {https://doi.org/10.1609/aaai.v33i01.3301346},
	doi          = {10.1609/AAAI.V33I01.3301346},
	bibsource    = {dblp computer science bibliography, https://dblp.org}
}

@article{TEA,
	author       = {Zijian Li and
	Ruichu Cai and
	Fengzhu Wu and
	Sili Zhang and
	Hao Gu and
	Yuexing Hao and
	Yuguang Yan},
	title        = {{TEA:} {A} Sequential Recommendation Framework via Temporally Evolving
	Aggregations},
	journal      = {{IEEE} Trans. Neural Networks Learn. Syst.},
	volume       = {35},
	number       = {2},
	pages        = {2628--2639},
	year         = {2024},
	url          = {https://doi.org/10.1109/TNNLS.2022.3190534},
	doi          = {10.1109/TNNLS.2022.3190534},
	bibsource    = {dblp computer science bibliography, https://dblp.org}
}

@article{DGSR,
	author       = {Mengqi Zhang and
	Shu Wu and
	Xueli Yu and
	Qiang Liu and
	Liang Wang},
	title        = {Dynamic Graph Neural Networks for Sequential Recommendation},
	journal      = {{IEEE} Trans. Knowl. Data Eng.},
	volume       = {35},
	number       = {5},
	pages        = {4741--4753},
	year         = {2023},
	url          = {https://doi.org/10.1109/TKDE.2022.3151618},
	doi          = {10.1109/TKDE.2022.3151618},
	bibsource    = {dblp computer science bibliography, https://dblp.org}
}

@inproceedings{Redrec,
	author       = {Runfeng Yao and
	Weisheng Xu and
	Zhenyu Liu and
	Yang Wang and
	Zhen Li and
	Yuanyuan Qiao and
	Jie Yang},
	title        = {Redrec: Relation and Dynamic Aware Graph Convolutional Network for
	Sequential Recommendation},
	booktitle    = {8th {IEEE} International Conference on Network Intelligence and Digital
	Content, {IC-NIDC} 2023, Beijing, China, November 3-5, 2023},
	pages        = {192--196},
		year         = {2023},
	url          = {https://doi.org/10.1109/IC-NIDC59918.2023.10390573},
	doi          = {10.1109/IC-NIDC59918.2023.10390573},
	bibsource    = {dblp computer science bibliography, https://dblp.org}
}

@inproceedings{GRU4Rec,
	author       = {Bal{\'{a}}zs Hidasi and
	Alexandros Karatzoglou and
	Linas Baltrunas and
	Domonkos Tikk},
	editor       = {Yoshua Bengio and
	Yann LeCun},
	title        = {Session-based Recommendations with Recurrent Neural Networks},
	booktitle    = {4th International Conference on Learning Representations, {ICLR} 2016,
	San Juan, Puerto Rico, May 2-4, 2016, Conference Track Proceedings},
	year         = {2016},
	url          = {http://arxiv.org/abs/1511.06939},
	bibsource    = {dblp computer science bibliography, https://dblp.org}
}

@inproceedings{KrNN-P,
	author       = {Xiang Lin and
	Shuzi Niu and
	Yiqiao Wang and
	Yucheng Li},
	editor       = {Kevyn Collins{-}Thompson and
	Qiaozhu Mei and
	Brian D. Davison and
	Yiqun Liu and
	Emine Yilmaz},
	title        = {K-plet Recurrent Neural Networks for Sequential Recommendation},
	booktitle    = {The 41st International {ACM} {SIGIR} Conference on Research {\&}
	Development in Information Retrieval, {SIGIR} 2018, Ann Arbor, MI,
	USA, July 08-12, 2018},
	pages        = {1057--1060},
	year         = {2018},
	url          = {https://doi.org/10.1145/3209978.3210159},
	doi          = {10.1145/3209978.3210159},
	bibsource    = {dblp computer science bibliography, https://dblp.org}
}

@inproceedings{BERT4Rec,
	author       = {Fei Sun and
	Jun Liu and
	Jian Wu and
	Changhua Pei and
	Xiao Lin and
	Wenwu Ou and
	Peng Jiang},
	editor       = {Wenwu Zhu and
	Dacheng Tao and
	Xueqi Cheng and
	Peng Cui and
	Elke A. Rundensteiner and
	David Carmel and
	Qi He and
	Jeffrey Xu Yu},
	title        = {BERT4Rec: Sequential Recommendation with Bidirectional Encoder Representations
	from Transformer},
	booktitle    = {Proceedings of the 28th {ACM} International Conference on Information
	and Knowledge Management, {CIKM} 2019, Beijing, China, November 3-7,
	2019},
	pages        = {1441--1450},
	year         = {2019},
	url          = {https://doi.org/10.1145/3357384.3357895},
	doi          = {10.1145/3357384.3357895},
	bibsource    = {dblp computer science bibliography, https://dblp.org}
}

@inproceedings{CLSR,
	author       = {Yu Zheng and
	Chen Gao and
	Jianxin Chang and
	Yanan Niu and
	Yang Song and
	Depeng Jin and
	Yong Li},
	editor       = {Fr{\'{e}}d{\'{e}}rique Laforest and
	Rapha{\"{e}}l Troncy and
	Elena Simperl and
	Deepak Agarwal and
	Aristides Gionis and
	Ivan Herman and
	Lionel M{\'{e}}dini},
	title        = {Disentangling Long and Short-Term Interests for Recommendation},
	booktitle    = {{WWW} '22: The {ACM} Web Conference 2022, Virtual Event, Lyon, France,
	April 25 - 29, 2022},
	pages        = {2256--2267},
	year         = {2022},
	url          = {https://doi.org/10.1145/3485447.3512098},
	doi          = {10.1145/3485447.3512098},
	bibsource    = {dblp computer science bibliography, https://dblp.org}
}

@article{wu2022survey,
	author       = {Le Wu and
	Xiangnan He and
	Xiang Wang and
	Kun Zhang and
	Meng Wang},
	title        = {A Survey on Accuracy-Oriented Neural Recommendation: From Collaborative
	Filtering to Information-Rich Recommendation},
	journal      = {{IEEE} Trans. Knowl. Data Eng.},
	volume       = {35},
	number       = {5},
	pages        = {4425--4445},
	year         = {2023},
	url          = {https://doi.org/10.1109/TKDE.2022.3145690},
	doi          = {10.1109/TKDE.2022.3145690},
	bibsource    = {dblp computer science bibliography, https://dblp.org}
}

@article{liao2022heterogeneous,
	author       = {Wenhui Liao and
	Qian Zhang and
	Bo Yuan and
	Guangquan Zhang and
	Jie Lu},
	title        = {Heterogeneous Multidomain Recommender System Through Adversarial Learning},
	journal      = {{IEEE} Trans. Neural Networks Learn. Syst.},
	volume       = {34},
	number       = {11},
	pages        = {8965--8977},
	year         = {2023},
	url          = {https://doi.org/10.1109/TNNLS.2022.3154345},
	doi          = {10.1109/TNNLS.2022.3154345},
	bibsource    = {dblp computer science bibliography, https://dblp.org}
}

@article{zheng2021incorporating,
	author       = {Yu Zheng and
	Chen Gao and
	Xiangnan He and
	Depeng Jin and
	Yong Li},
	title        = {Incorporating Price into Recommendation With Graph Convolutional Networks},
	journal      = {{IEEE} Trans. Knowl. Data Eng.},
	volume       = {35},
	number       = {2},
	pages        = {1609--1623},
	year         = {2023},
	url          = {https://doi.org/10.1109/TKDE.2021.3091160},
	doi          = {10.1109/TKDE.2021.3091160},
	bibsource    = {dblp computer science bibliography, https://dblp.org}
}

@article{CF,
  author       = {Xiangnan He and
Lizi Liao and
Hanwang Zhang and
Liqiang Nie and
Xia Hu and
Tat{-}Seng Chua},
editor       = {Rick Barrett and
Rick Cummings and
Eugene Agichtein and
Evgeniy Gabrilovich},
title        = {Neural Collaborative Filtering},
booktitle    = {Proceedings of the 26th International Conference on World Wide Web,
{WWW} 2017, Perth, Australia, April 3-7, 2017},
pages        = {173--182},
publisher    = {{ACM}},
year         = {2017},
url          = {https://doi.org/10.1145/3038912.3052569},
doi          = {10.1145/3038912.3052569},
bibsource    = {dblp computer science bibliography, https://dblp.org}
}

@article{MF,
	author       = {Yehuda Koren and
	Robert M. Bell and
	Chris Volinsky},
	title        = {Matrix Factorization Techniques for Recommender Systems},
	journal      = {Computer},
	volume       = {42},
	number       = {8},
	pages        = {30--37},
	year         = {2009},
	url          = {https://doi.org/10.1109/MC.2009.263},
	doi          = {10.1109/MC.2009.263},
	bibsource    = {dblp computer science bibliography, https://dblp.org}
}

@inproceedings{Markov-chains,
	author       = {Steffen Rendle and
	Christoph Freudenthaler and
	Lars Schmidt{-}Thieme},
	editor       = {Michael Rappa and
	Paul Jones and
	Juliana Freire and
	Soumen Chakrabarti},
	title        = {Factorizing personalized Markov chains for next-basket recommendation},
	booktitle    = {Proceedings of the 19th International Conference on World Wide Web,
	{WWW} 2010, Raleigh, North Carolina, USA, April 26-30, 2010},
	pages        = {811--820},
	year         = {2010},
	url          = {https://doi.org/10.1145/1772690.1772773},
	doi          = {10.1145/1772690.1772773},
	bibsource    = {dblp computer science bibliography, https://dblp.org}
}

@inproceedings{Unbiased2020,
	author       = {Yuta Saito and
	Suguru Yaginuma and
	Yuta Nishino and
	Hayato Sakata and
	Kazuhide Nakata},
	editor       = {James Caverlee and
	Xia (Ben) Hu and
	Mounia Lalmas and
	Wei Wang},
	title        = {Unbiased Recommender Learning from Missing-Not-At-Random Implicit
	Feedback},
	booktitle    = {{WSDM} '20: The Thirteenth {ACM} International Conference on Web Search
	and Data Mining, Houston, TX, USA, February 3-7, 2020},
	pages        = {501--509},
	year         = {2020},
	url          = {https://doi.org/10.1145/3336191.3371783},
	doi          = {10.1145/3336191.3371783},
	bibsource    = {dblp computer science bibliography, https://dblp.org}
}

@inproceedings{VIB-GSL,
	author       = {Qingyun Sun and
	Jianxin Li and
	Hao Peng and
	Jia Wu and
	Xingcheng Fu and
	Cheng Ji and
	Philip S. Yu},
	title        = {Graph Structure Learning with Variational Information Bottleneck},
	booktitle    = {Thirty-Sixth {AAAI} Conference on Artificial Intelligence, {AAAI}
	2022, Thirty-Fourth Conference on Innovative Applications of Artificial
	Intelligence, {IAAI} 2022, The Twelveth Symposium on Educational Advances
	in Artificial Intelligence, {EAAI} 2022 Virtual Event, February 22
	- March 1, 2022},
	pages        = {4165--4174},
	year         = {2022},
	url          = {https://doi.org/10.1609/aaai.v36i4.20335},
	doi          = {10.1609/AAAI.V36I4.20335},
	bibsource    = {dblp computer science bibliography, https://dblp.org}
}

@article{LLM4DSR,
	author       = {Bohao Wang and
	Feng Liu and
	Jiawei Chen and
	Yudi Wu and
	Xingyu Lou and
	Jun Wang and
	Yan Feng and
	Chun Chen and
	Can Wang},
	title        = {{LLM4DSR:} Leveraing Large Language Model for Denoising Sequential
	Recommendation},
	journal      = {CoRR},
	volume       = {abs/2408.08208},
	year         = {2024},
	url          = {https://doi.org/10.48550/arXiv.2408.08208},
	doi          = {10.48550/ARXIV.2408.08208},
	eprinttype    = {arXiv},
	eprint       = {2408.08208},
	bibsource    = {dblp computer science bibliography, https://dblp.org}
}

@article{NiDen,
	author       = {Haibo Ye and
	Lijun Zhang and
	Yuan Yao and
	Sheng{-}Jun Huang},
	title        = {Denoised Graph Collaborative Filtering via Neighborhood Similarity
	and Dynamic Thresholding},
	journal      = {{IEEE} Trans. Big Data},
	volume       = {10},
	number       = {6},
	pages        = {683--693},
	year         = {2024},
	url          = {https://doi.org/10.1109/TBDATA.2024.3453765},
	doi          = {10.1109/TBDATA.2024.3453765},
	bibsource    = {dblp computer science bibliography, https://dblp.org}
}

@inproceedings{UDT,
	author       = {Haoyan Chua and
	Yingpeng Du and
	Zhu Sun and
	Ziyan Wang and
	Jie Zhang and
	Yew{-}Soon Ong},
	editor       = {Tommaso Di Noia and
	Pasquale Lops and
	Thorsten Joachims and
	Katrien Verbert and
	Pablo Castells and
	Zhenhua Dong and
	Ben London},
	title        = {Unified Denoising Training for Recommendation},
	booktitle    = {Proceedings of the 18th {ACM} Conference on Recommender Systems, RecSys
	2024, Bari, Italy, October 14-18, 2024},
	pages        = {612--621},
	year         = {2024},
	url          = {https://doi.org/10.1145/3640457.3688109},
	doi          = {10.1145/3640457.3688109},
	bibsource    = {dblp computer science bibliography, https://dblp.org}
}

@article{SubGCL,
	author       = {Yi Yang and
	Shaopeng Guan and
	Xiaoyang Wen},
	title        = {Enhancing robustness in implicit feedback recommender systems with
	subgraph contrastive learning},
	journal      = {Inf. Process. Manag.},
	volume       = {62},
	number       = {3},
	pages        = {103962},
	year         = {2025},
	url          = {https://doi.org/10.1016/j.ipm.2024.103962},
	doi          = {10.1016/J.IPM.2024.103962},
	bibsource    = {dblp computer science bibliography, https://dblp.org}
}

@article{HetroFair,
  author       = {Nemat Gholinejad and
                  Mostafa Haghir Chehreghani},
  title        = {Heterophily-aware fair recommendation using graph convolutional networks},
  journal      = {Neurocomputing},
  volume       = {661},
  pages        = {131956},
  year         = {2026},
  url          = {https://doi.org/10.1016/j.neucom.2025.131956},
  doi          = {10.1016/J.NEUCOM.2025.131956},
  bibsource    = {dblp computer science bibliography, https://dblp.org}
}

@article{zhou2020graph,
	author       = {Jie Zhou and
	Ganqu Cui and
	Shengding Hu and
	Zhengyan Zhang and
	Cheng Yang and
	Zhiyuan Liu and
	Lifeng Wang and
	Changcheng Li and
	Maosong Sun},
	title        = {Graph neural networks: {A} review of methods and applications},
	journal      = {{AI} Open},
	volume       = {1},
	pages        = {57--81},
	year         = {2020},
	url          = {https://doi.org/10.1016/j.aiopen.2021.01.001},
	doi          = {10.1016/J.AIOPEN.2021.01.001},
	bibsource    = {dblp computer science bibliography, https://dblp.org}
}

@article{gao2023survey,
	author       = {Chen Gao and
	Yu Zheng and
	Nian Li and
	Yinfeng Li and
	Yingrong Qin and
	Jinghua Piao and
	Yuhan Quan and
	Jianxin Chang and
	Depeng Jin and
	Xiangnan He and
	Yong Li},
	title        = {A Survey of Graph Neural Networks for Recommender Systems: Challenges,
	Methods, and Directions},
	journal      = {Trans. Recomm. Syst.},
	volume       = {1},
	number       = {1},
	pages        = {1--51},
	year         = {2023},
	url          = {https://doi.org/10.1145/3568022},
	doi          = {10.1145/3568022},
	bibsource    = {dblp computer science bibliography, https://dblp.org}
}

@inproceedings{10.1145/3637528.3671692,
	author = {He, Zhuangzhuang and Wang, Yifan and Yang, Yonghui and Sun, Peijie and Wu, Le and Bai, Haoyue and Gong, Jinqi and Hong, Richang and Zhang, Min},
	title = {Double Correction Framework for Denoising Recommendation},
	year = {2024},
	isbn = {9798400704901},
	publisher = {Association for Computing Machinery},
	address = {New York, NY, USA},
	url = {https://doi.org/10.1145/3637528.3671692},
	doi = {10.1145/3637528.3671692},
	booktitle = {Proceedings of the 30th ACM SIGKDD Conference on Knowledge Discovery and Data Mining},
	pages = {1062–1072},
	numpages = {11},
	location = {Barcelona, Spain},
	series = {KDD '24}
}

\end{document}